\documentclass[useAMS,usenatbib]{mn2e}

\usepackage{epsfig}
\usepackage{textcomp}
\usepackage{longtable}
\usepackage{lscape}

\newcommand{\xmm} {{\it XMM-Newton}}
\newcommand{\cmsq} {cm$^{-2}$}
\newcommand{\nh} {$N_{\rm{H}}$}

\newcommand{\mic}{{${\umu}$m}}
\newcommand{\oiii}{{\rm{[O\,\sc{iii}]}}}
\newcommand{\hii}{{\rm{H\,\sc{ii}}}}

\newcommand{\ergs}{\mbox{\thinspace erg\thinspace s$^{-1}$}}

\title[XMM survey of the 12MGS]{An XMM-Newton spectral survey of 12 micron selected galaxies. II. Implications for AGN selection and unification}

\author[M. Brightman and K. Nandra]{Murray Brightman$^{2,1}\thanks{E-mail: mbright@mpe.mpg.de}$ and Kirpal Nandra$^{2,1}$\\
$^{1}$Astrophysics Group, Imperial College London, Blackett Laboratory, Prince Consort Road, London SW7 2AW\\
$^{2}$Max-Planck-Institut f\"{u}r extraterrestrische Physik, Giessenbachstrasse 1, D-85748, Garching bei M\"{u}nchen, Germany\\}

\begin{document}

\date{Accepted 0000 December 00. Received 0000 December 00; in original form 0000 October 00}

\pagerange{\pageref{firstpage}--\pageref{lastpage}} \pubyear{0000}

\maketitle

\label{firstpage}

\begin{abstract}

We present a multi-waveband analysis of a 126 galaxy sub-sample of the 12 micron galaxy sample (12MGS), for which we have carried out a detailed X-ray spectral analysis of in a previous paper. We determine the activity class of the galaxies by way of optical line ratio diagnostics and we characterise the optical classes by their X-ray, 12 \mic\ and \oiii\ luminosities and by their X-ray spectral properties. Our most interesting results from this investigation are as follows: (i) Seyfert (Sy) 1s and Sy 2s show a significantly different X-ray luminosity distributions from each other (ii) Sy 2 galaxies with a detection of an HBLR show a significantly higher X-ray luminosity than those without a detection, supporting the findings of \citet{tran03} (iii) Sy 1s also present a significantly different 12 \mic\ luminosity distribution from both intermediate Sy types and Sy 2s (iv) the Seyfert 2 fraction decreases towards high X-ray luminosities (v) X-ray indications of AGN power agree well with the optical classifications (vi) There is X-ray evidence for the presence of an AGN in 17\% of \hii/AGN composite galaxies and 40\% of LINERs (vii) we advocate the use of a 2-10 keV X-ray luminosity of 10$^{41}$ \ergs\ in the X-ray selection of AGN, rather than 10$^{42}$ \ergs, which we find gives a contamination rate of only 3\% from star-forming galaxies. (viii) from an analysis of the X-ray power-law index, $\Gamma$, we find that Sy 1s and Sy 2 have the same intrinsic distributions, implying that the central engines are the same, in support of AGN unification schemes (ix) in 24\% of cases the absorption measured in the X-ray spectra does not correspond directly with that implied in the optical band from the visibility of the broad line regions (BLRs), which is in conflict with AGN unification schemes (x) We confirm previous work showing that the obscured fraction in AGN declines at high X-ray luminosity, but also find a decrease at low luminosity having peaked at $L_{\rm X}\sim10^{42}$ \ergs, suggesting that source luminosity has a large effect on the obscuring material, therefore also calling for a modification to unified schemes (xi) The average obscured and Compton thick fractions for this sample are 62 $\pm$ 5\% and 20 $\pm$ 4\% respectively, which are higher than hard X-ray and optically selected samples, therefore supporting mid-infrared (MIR) selection as a relatively unbiased method of selecting AGN (xii) we assess the use of the `T' ratio ($F_{\rm X}/F_{\rm [OIII]}$) for selecting Compton thick candidates. We conclude here that this quantity can often be unreliable due uncertainties in the extinction corrections to the \oiii\ flux. These results have important impacts on AGN selection and unification and the results from the 12MGS are particularly useful as a local analogue to {\it Spitzer}/MIPS 24 \mic\ samples selected at z=1, as observed 24 \mic\ emission originates at rest-frame 12 \mic\ in sources at this redshift.

\end{abstract}


\section{Introduction}

There are currently ongoing debates in astrophysics concerning many aspects of nuclear activity in galaxies. It has become widely accepted though, that most massive galaxies harbour a super-massive black hole in their nuclei \citep{kormendy88, magorrian98}, and that this component is central to the growth and evolution of the host galaxy itself \citep[eg. ][]{ferrarese00,gebhardt00}. It is believed that accretion of matter onto these black holes is the energy source for what we know as active galactic nuclei (AGN). The term `active' when applied to the nucleus of a galaxy is historically quite general and describes a nucleus which displays characteristics which cannot be attributed to normal stellar processes. This can be highly wavelength dependent and as such, the identification of AGN depends on the wavelength regime used. A prime example is NGC 6240, which was identified as an AGN from its X-ray luminosity, which exceeds any X-ray luminosity observed from pure star-forming galaxies \citep{schultz98}, but does not appear as an AGN in the optical. These types of AGN, which do not show themselves at all wavelengths, can be called `hidden' AGN.  AGN activity continues to be uncovered in previously thought of normal galaxies, such as with the use of MIR high excitation lines \citep[e.g.][]{goulding09} or high spatial resolution X-ray imaging \citep[e.g.][]{grier10}. It is thus important to investigate the identification technique which selects AGN most completely.

AGN also appear in many different types, ranging from low ionisation nuclear emission-line regions \citep[LINERs,][]{heckman80} to quasars, and attempts have been made to unify these types into a single scheme. The most successful unification scheme explains the difference between the different Seyfert types by the orientation of the observer with respect to a circumnuclear structure of dust, now commonly believed to be a torus \citep{antonucci93, urry95}. This scheme has held up well, owing to it simplicity, but has come under increasing pressure of late. For example, there is increasing evidence and supporting theoretical work to suggest that at low luminosity and/or accretion rate, AGN appear differently from their higher luminosity counterparts \citep[e.g.][]{laor03, nicastro03, tran03, elitzur06, hopkins09}. LINERs may also be accounted for in part by these low luminosity AGN models, but a concensus on the power generation mechanism in LINERs has not yet been achieved. Obscuration in AGN may also be dependent on luminosity, as has been found in several studies at high luminosity \citep[e.g.][]{ueda03}, but also suggested at low luminosity \citep{akylas09, zhang09, burlon10}. This has important consequences for the current AGN unification paradigm, which attributes the sole difference between different AGN types to the orientation with respect to the observer.

%There are currently ongoing debates in astrophysics concerning many aspects of nuclear activity in galaxies. These include the nature of low luminosity active galactic nuclei \citep[LLAGNs,][]{ho97}, low ionisation nuclear emission-line regions \citep[LINERs,][]{heckman80} and composite AGN/star-forming galaxies \citep{kewley01}; the nature of intermediate Seyfert type galaxies \citep[Seyfert 1.2, 1.5, 1.8 and 1.9s,][]{winkler92}, as well as the nature of obscuration in AGN and the unification schemes \citep{antonucci93, urry95}. 

%There is increasing evidence and supporting theoretical work to suggest that at low luminosity and/or accretion rate, AGN appear differently from their higher luminosity counterparts \citep[e.g.][]{laor03, nicastro03, tran03, elitzur06, hopkins09}. LINERs may be accounted for in part by these LLAGN models, but a concensus on the power generation mechanism in LINERs has not yet been achieved. Obscuration in AGN may also be dependent on luminosity, as has been found in several studies at high luminosity \citep[e.g.][]{ueda03}, but also suggested at low luminosity \citep{akylas09}. This has important consequences for the current AGN unification paradigm, which attributes the sole difference between different AGN types to the orientation of the observer.

To investigate the issues of AGN selection and unification, large, statistically complete galaxy samples are required. Most of these are based on flux limited surveys, for example the Revised Shapley-Ames Catalogue of Bright Galaxies \citep[RSA,][]{sandage79}  or the CfA sample \citep{huchra92}, both selected in the optical. Optical selection, however, is largely affected by extinction, presenting a bias against reddened sources. Galaxy samples selected at wavelengths less affected by extinction are thus optimal. Hard X-ray ($>$10 keV) surveys are ideal for avoiding biases against obscuration as at these wavelengths only the heaviest obscuration attenuates them. Current sensitivity at these wavelengths provided by the {\it Swift/BAT} and {\it INTEGRAL} surveys produces samples of sources down to $L_{\rm HX}\sim10^{41}$ \ergs\ \citep[e.g.][]{tueller08, beckmann09, tueller10, cusumano10}, and thus do not contain the lowest luminosity systems. 

An alternative wavelength for AGN selection is the MIR, where the primary radiation of the AGN is re-emitted after having been reprocessed by hot dust. The extended {\it IRAS} 12 micron galaxy sample \citep*[12MGS -][RMS]{rush93} is a sample of 893 MIR selected local galaxies which contains a high fraction of AGN (13\% at the time of publication, RMS). The sample is taken from the {\it IRAS Faint Source Catalogue, version 2 (FSC-2)} and imposes a flux limit of 0.22 Jy, including only sources with a rising flux density from 12 to 60 microns (to exclude stars) and with a galactic latitude of $|b| \geq$25 deg. Being selected in the mid-infrared (MIR) this sample is also relatively unbiased against absorption, low luminosity systems and `hidden AGN'. \citet{spinoglio89} showed that a wide variety of AGN types emit a constant fraction of their bolometric luminosities at this wavelength, and furthermore shown to be true for star forming galaxies as well by \citet{spinoglio95}. The 12MGS should therefore also be representative of the true number of different active galaxy types in relation to each other and thus ideal for population statistics.

Empirically, 12 micron selection appears to be relatively unbiased with respect to extinction. However, it is expected theoretically that 12 \mic\ emission is suppressed in heavily obscured AGN. In the radiation transfer models of \citet{pier92}, the authors investigate the infrared emission of centrally illuminated smooth dust tori, thought to exist in AGN. A main conclusion of this is that these dust tori do not emit isotropically in the mid-IR, and that emission in the `face-on' direction is greater than that in the `edge-on' direction. This has important consequences for our work, as it would suggest that 12 \mic\ selection is biased against these edge on systems. Alternatively, much support has been gained recently for a `clumpy' torus in AGN, where the dust distribution is not smooth, but instead made up of individual clouds \citep[e.g.][]{ralmeida09,hoenig10,mullaney11}. \citet{nenkova08} present a model describing the infra-red emission of such a torus and find that at 12 \mic, and at all IR wavelengths, the torus emission is isotropic, in contrast to the smooth torus models. A key observational test of this would be the ratio of the X-ray to mid-IR fluxes, which should be higher for heavily absorbed systems in smooth torus distributions, when absorption in the X-ray band has been accounted for, due to the higher mid-IR emission in the face-on systems.  The models of \citet{pier92} predict an order of magnitude difference in the 12 \mic\ flux for an increase in the \nh\ from $\sim10^{23}$ to $\sim10^{24}$ for a torus seen edge on. \citet{horst07}, however, show that Seyfert 1s, Seyfert 2 and even Compton thick AGN show the same tight correlation between their X-ray and mid-IR fluxes, which supports the clumpy torus model over the smooth one. They also rule out any contamination of the mid-IR flux from the host galaxy by using high resolution, adaptive optics assisted, 12.3 \mic\ imaging, isolating the torus spatially. These results are in support of 12 \mic\ being a relatively unbiased selection method for AGN.

In this paper, we aim to broadly investigate the nuclear activity in local ($z<0.1$) galaxies, using the 12MGS as a relatively unbiased and representative parent sample. We base this investigation on a sub-sample of 126 galaxies in the 12MGS which we presented in a previous publication \citep[][paper I hereafter]{brightman10}. This sub-sample consists of all galaxies in the 12MGS for which an \xmm\ observation was available as of December 2008 (which included serendipitous observations of NGC 0214, 4559 and 7771), and for which the X-ray spectrum could be at least fitted with the most basic spectral model. For more details regarding the X-ray observations and spectral analysis, the reader is referred to paper I. In paper I, we presented a detailed X-ray spectral analysis of these galaxies, assessing the intrinsic luminosity, $L_{\rm X}$, the photon index, $\Gamma$ and the line of sight absorption, \nh, which are key parameters for our investigation here. We start in Section \ref{optclass} where we uniformly classify the activity type in our sample using optical emission line data from the literature and BPT \citep*{baldwin81} diagnostics. Secondly in Section \ref{lumchar} we characterise the different optical types in 2-10 keV, 12 \mic\ and \oiii\ luminosities and in Section \ref{xrayprop} we investigate the X-ray properties of these types, including the continuum slope, $\Gamma$ and the obscuration, \nh. Finally in Section \ref{xrayirrel} we investigate the relationship between the X-ray and MIR emission in this sample. We discuss these results within the context of AGN selection and unification in Section \ref{discussion} and present our conclusions in Section \ref{conclusions}.

\section{Optical emission line activity classification}
\label{optclass}

In paper I we used an observed 2-10 keV X-ray luminosity of 10$^{42}$ \ergs\ to identify unambiguous AGN activity in our sample. This method is clearly very crude as it will miss all low luminosity AGN. The classical method of defining activity type is via so called BPT  diagrams which use the ratios of optical emission lines to determine the dominant source of ionising radiation of emission line galaxies, be it photo-ionisation by stars, by a harder non-thermal source such as an AGN or by collisional excitation by shocks, as may be the case in LINERs. Here we apply such a scheme in order to investigate the X-ray properties of the various optical types.

\begin{figure*}
\begin{minipage}{180mm}
\includegraphics[width=180mm]{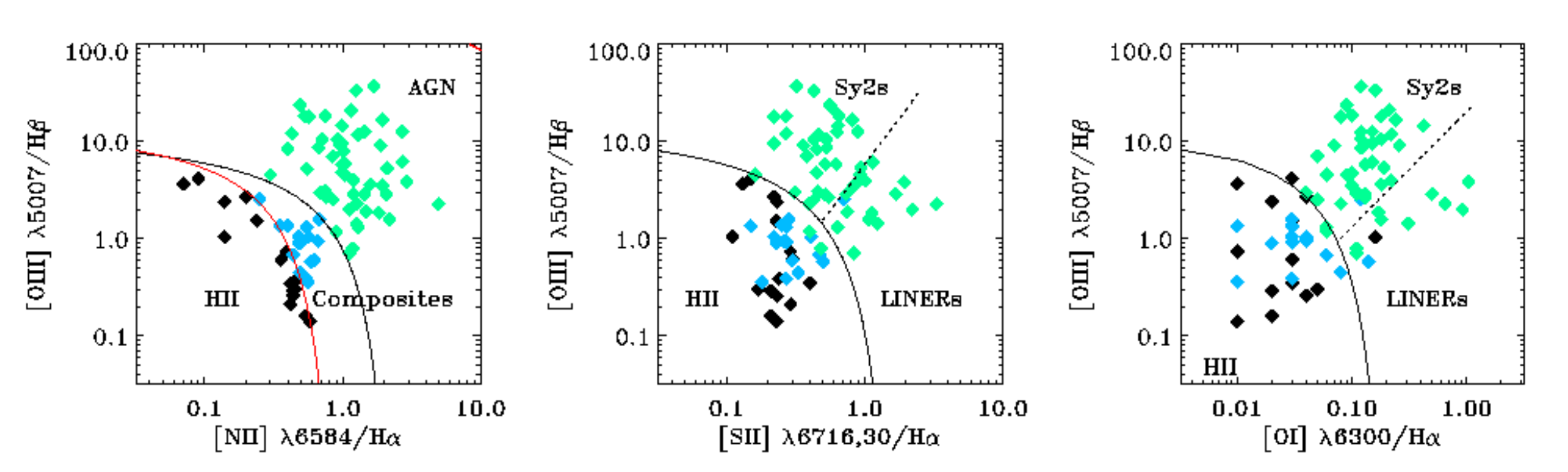}
\caption{BPT diagrams showing diagram (1) on the left using \oiii/H$\beta$ vs.[N {\sc ii}]/H$\alpha$, diagram (2) in the middle using \oiii/H$\beta$ vs. [S {\sc ii}]/H$\alpha$ and diagram (3) on the right using \oiii/H$\beta$ vs. [O {\sc i}]/H$\alpha$. \hii\ galaxies (black), Seyfert 2s (blue) and LINERs (green) are separated using the classification scheme of \citet{kewley06}. Diagram 1 identifies pure \hii\ galaxies (black), \hii/AGN composites (light blue) and pure AGN (turquoise), but does not distinguish between Sy 2s and LINERs. We plot the types classified in diagram 1 on the following two diagrams to display the agreement between the two.}
\label{fig:mwbanalysis_bpt}
\end{minipage}
\end{figure*}

We use the scheme of \citet{kewley06} to classify the galaxies in our sample from optical narrow line emission (Fig. \ref {fig:mwbanalysis_bpt}). This scheme uses theoretical `maximal starburst' lines from \citet{kewley01} which are derived from stellar population and photoionisation models and define the limit of `pure' stellar photoionisation. Any sources which lie above these lines are likely to be photoionized by another source of ionization such as an AGN or shocks. This scheme relies on three diagrams, diagram (1) using \oiii/H$\beta$ vs. [N {\sc ii}]/H$\alpha$, diagram (2) using \oiii/H$\beta$ vs. [S {\sc ii}]/H$\alpha$ and diagram (3) using \oiii/H$\beta$ vs. [O {\sc i}]/H$\alpha$. They also define a region of composite `starburst/AGN' activity on diagram 1, which lies below the theoretical starburst limit, but above an empirically defined pure-starburst limit \citep{kauffmann03}. They argue that since the [N {\sc ii}]/H$\alpha$ ratio is most sensitive to the presence of an AGN due to the ratio `saturating' at high metalicities, which exist in extreme \hii\ galaxies, any source beyond the pure starburst limit must contain an AGN. On diagrams 2 and 3, they derive new empirical separation lines between Seyfert 2s and LINERs using high signal-to-noise SDSS data, which show distinct branches on these diagrams, belonging to these emission line galaxy types. 

We have obtained narrow line ratios from a compilation of unpublished data from Rodriguez et al. (in preparation) for 70 of the galaxies in our sample of 126 sources. For the rest, where they exist, we have compiled from the literature the \oiii/H$\beta$, [O {\sc i}]/H$\alpha$, [N {\sc ii}]/H$\alpha$ and [S {\sc ii}]/H$\alpha$ line ratios which are needed for this classification method.  In five cases we have used SDSS DR7 emission line data, where no literature line ratio data exists. These data originate from the {\sc specBS} code which extracts emission line data from the {\sc spectro2D} pipeline reduced spectra\footnote{http://www.sdss.org/dr7/products/spectra/index.html} done at Princeton University. Table \ref{table_optdat} presents these data, including the reference for the line ratio data. Fig. \ref{fig:mwbanalysis_bpt} presents the BPT classifications diagrammatically.

We adopt a classification based on the three diagrams. In the cases where the classification differs on one or two diagrams for the same galaxy, we use a composite classification such as `\hii/AGN' , `Sy2/LINER' or `\hii/LINER'. If the classification differs between diagram 2 and diagram 3, we choose the classification given by diagram 3, as [S {\sc ii}]/H$\alpha$ is said to be the least effective diagnostic in activity classification and [O {\sc i}]/H$\alpha$ is most sensitive to shocks, making it better at identifying LINERs \citep{kewley01}. Each diagram requires 4 emission lines for the classification. If data for all 4 lines are not available, we cannot use that diagram for the classification. 

Finally we supersede this activity classification, which is based on narrow line diagnostics, if a classification of a Seyfert 1-1.9 has been given in the literature. The classification of these sources based on the detection of broad lines is dependent on the methods used by each author, and thus differing classifications can be given for one source. It is also possible that a source changes classification. We have attempted to be as uniform as possible with our literature choice, taking the majority of the broad line classifications from Rodriguez, et al (25/33), and \citet{degrijp92} (4/33).

%by Rodriguez, et al or} in the literature based on the detection of broad components of the H$\alpha$ or H$\beta$ lines. {\bf The classification of these sources based on broad lines is dependent on the methods used by each author, and thus differing classifications can be given for one source. We take the majority (25/33) of the broad line classifications from Rodriguez, et al, and thus our classifications of these sources are close to being uniform.} 
%Table \ref{table_intro_syclass} gives the quantitative classification scheme for these objects.

 The results of this classification are also presented in Table \ref{table_optdat}. Our sample then consists of 12 Seyfert 1 galaxies, 16 Seyfert 1.2-1.5 galaxies, 5 Seyfert 1.8-1.9 galaxies (21 Seyfert 1.2-1.9 galaxies), 37 Seyfert 2 galaxies, 11 LINERs, 2 ambiguous Seyfert 2/LINERs, 13 \hii\ galaxies, 18 \hii/AGN composites, 3 ambiguous \hii/LINERs, and 9 galaxies which remain unclassified due to insufficient optical emission line data existing in the literature. 
 
The 12MGS as published by RMS contained 893 galaxies. At the time of publication these included 53 Seyfert 1s, 63 Seyfert 2s, 29 LINERs and 38 star forming galaxies, whereas the rest of the sample were called normal galaxies.  Table \ref{table:mwbanalysis_proportion} lists the proportions of the various optical types in the complete 12MGS and in our \xmm\ sub-sample in percentage form with their binomial errors. The ratio of Seyfert 2s to Seyfert 1 in the original sample was 1.19. If we group Seyfert 1, 1.2 and 1.5s together as Seyfert 1s and Seyfert 1.8, 1.9 and 2s together as Seyfert 2s as is commonly done, we have a ratio of 1.5 in our sub-sample. We also note, however, a ratio of 1.12 narrow line (Sy 2) to broad line (Sy 1-1.9) in our sample. The ratio of LINERs to Seyfert galaxies in the original sample is 0.25, whereas our sample contains 0.14 LINERs to Seyferts. And finally, the ratio of star-forming galaxies to Seyferts in the original sample was 0.33, whereas our sample contains 0.44 when \hii/AGN composite galaxies are included, or 0.19 when only pure star-forming galaxies are counted. We show here that the \xmm\ subsample we use is thus representative of the complete 12MGS as published by RMS, and not biased towards AGN. We do note however, that since RMS, optical type proportions have changed, for example, by the discovery of many Seyfert 2 galaxies in normal/starforming galaxies \citep[eg.][]{dopita98, thean01}. Furthermore, the 12MGS is not spectroscopically complete \citep{huntmalkan99}, so there may yet be undiscovered active galaxies within the normal galaxy population. However, the majority of sources not spectroscopically identified are southern galaxies, so we posit that the incompleteness is not caused by a bias against faint sources, rather by observability. The classification of the entire sample has not been done uniformly with a single classification scheme, but this will be the subject of future work by Rodriguez et al. (in preparation). In this study, we benefit from a uniform classification of all the galaxies in our sub-sample, which is almost spectroscopically complete.

\setcounter{table}{2}
\begin{table}
\centering
\caption{Proportion of optical types in our \xmm\ sub-sample compared with the proportions from the entire 12MGS \citep*[RMS,][]{rush93} with binomial errors. `all galaxies' refers to all 893 members of the 12MGS, `active galaxies' refers to all Seyferts, LINERs and \hii\ galaxies and `comp' refers to \hii/AGN composites.}
\label{table:mwbanalysis_proportion}
\begin{center}
\begin{tabular}{l l l}
\hline
Proportion		& 12MGS (RMS) 	& \xmm \\
 & & sub-sample \\
\hline
Sy/all galaxies		& 13 $\pm$ 1.1\%	& - \\
Sy/active galaxies	& 63 $\pm$ 3.1\%	& 56 $\pm$ 4.4\% \\
Sy 1.8-2/Sy		& 54 $\pm$ 4.6\%	& 60 $\pm$ 5.9\% \\
Sy 2/Sy			& - 				& 53 $\pm$ 5.9\% \\
LINER/(LINER+Sy)	& 20 $\pm$ 3.3\%	& 14 $\pm$ 3.9\% \\
\hii/(\hii+Sy)		& 25 $\pm$ 3.5\%	& 16 $\pm$ 4.0\% \\
(\hii+comp)/(\hii+comp+Sy) & - & 31 $\pm$ 4.6\% \\

\hline
\end{tabular}
\end{center}
\end{table}

In addition to the optical line ratios used for diagnostics, we also gather \oiii$\lambda$5007 line fluxes, which are often used as an indicator of the intrinsic luminosity of the central engine. We apply reddening corrections to these fluxes using the method described by \citet{veilleux87} which makes use of the observed Balmer decrement, H$\alpha$/H$\beta$ and assumes an intrinsic value of H$\alpha$/H$\beta$=2.85 for \hii\ like galaxies, and H$\alpha$/H$\beta$=3.1 for AGN. For Balmer decrements less than the assumed intrinsic value, we do not apply the reddening correction. These data are also presented in Table \ref{table_optdat}.

Having used optical emission line diagnostics to classify the galaxies in this sample, we can then define a set of AGN within the sample based on optical data. We call any galaxy with a Seyfert classification an AGN, of which there are 70 in our sample. If a hidden BLR (HBLR) has been detected, either using near-IR spectroscopy, or optical spectropolarimetry, we classify this as an AGN, regardless of the classification based on non-polarimetric optical spectroscopy. Eight  objects without a Seyfert classification, but with an HBLR are in our sample. We obtain this information from \citet{veroncetty06}, who present a compilation of HBLR data which has been compiled from the literature for their catalogue of AGN, which we give in Table \ref{table_tab2}. Table \ref{table_tab2} also lists all mulitwavelength data used in the analysis of this paper. 

\section{Luminosity characteristics of the 12MGS}
\label{lumchar}

\begin{figure*}
\begin{minipage}{180mm}
\includegraphics[width=180mm]{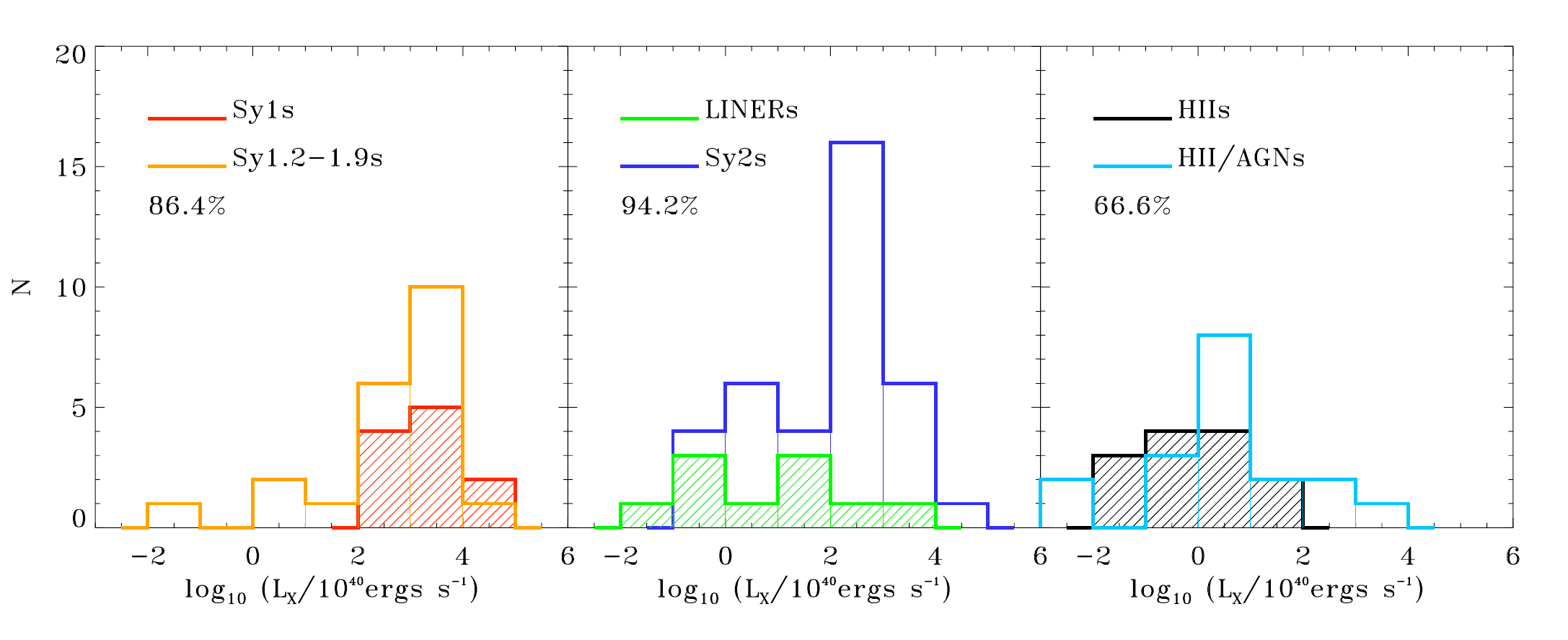}
\includegraphics[width=180mm]{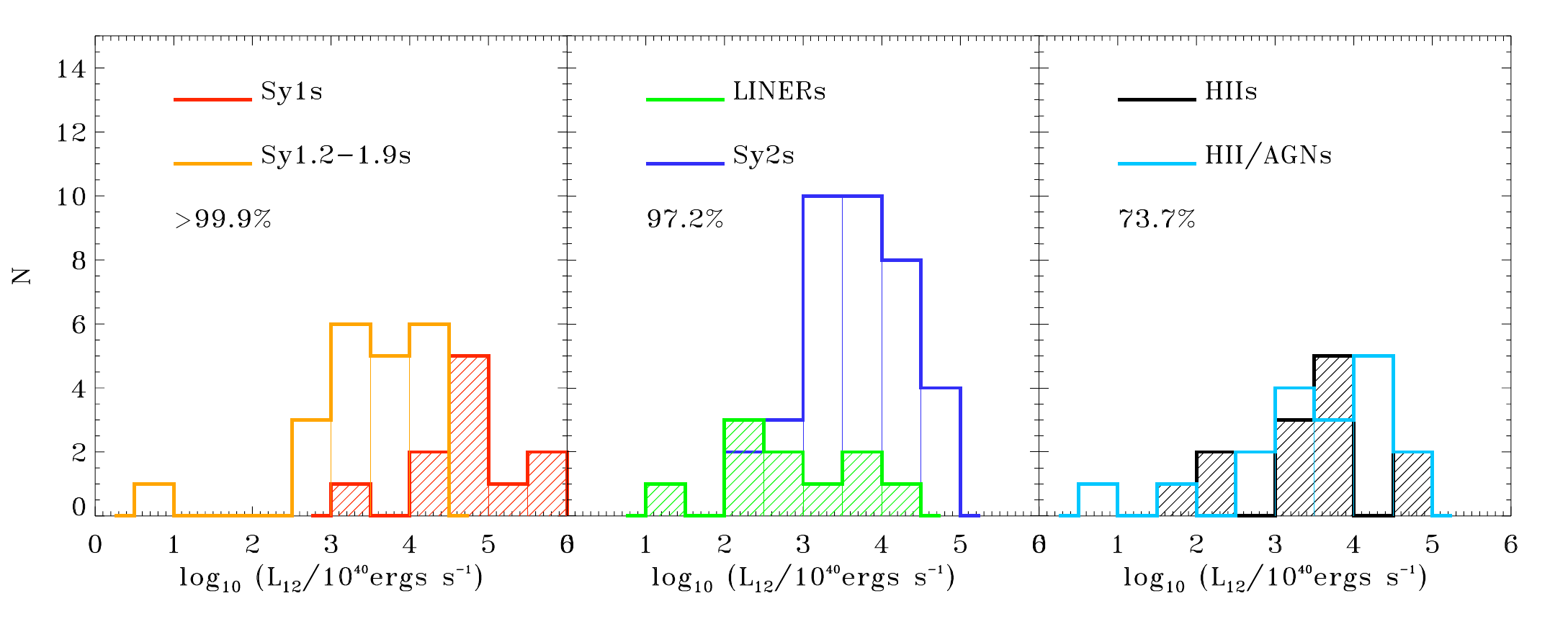}
\includegraphics[width=180mm]{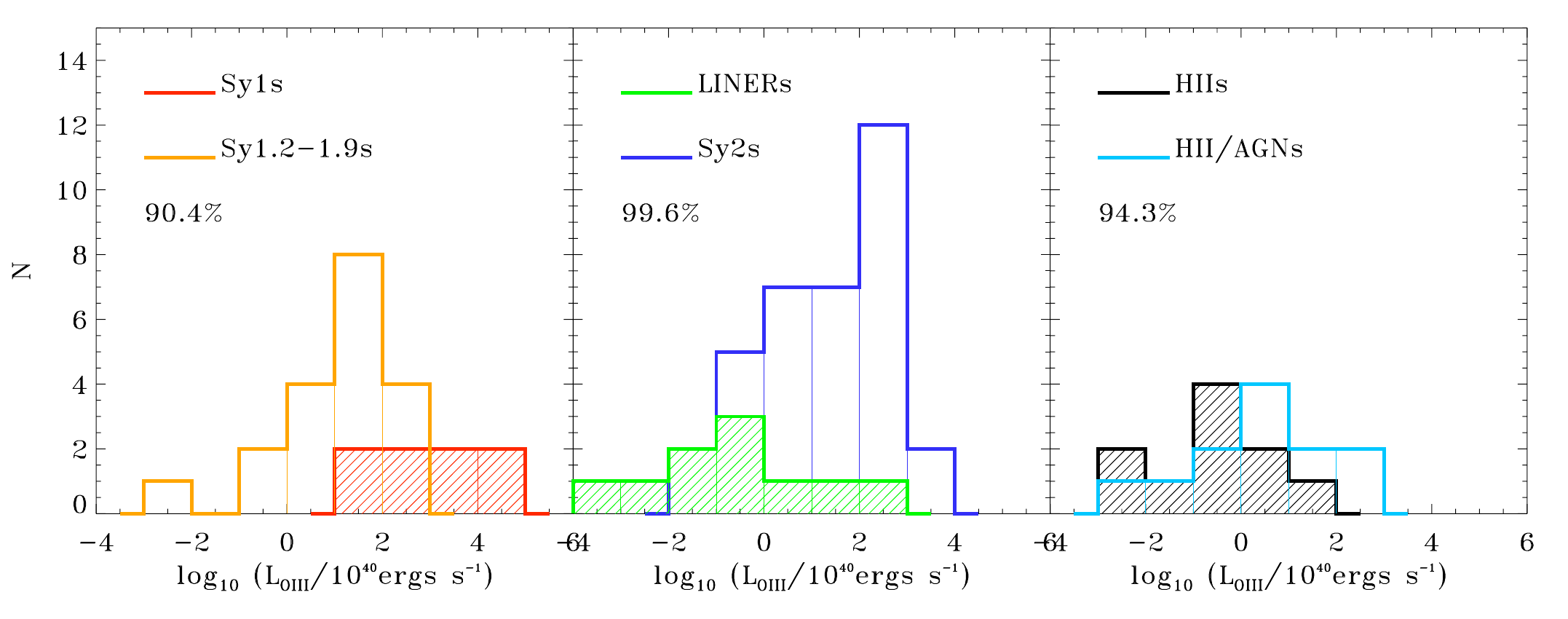}
 \caption{Luminosity distribution of the sources by optical type. Left panels - Seyfert 1s (red) and Seyfert 1.2-1.9 galaxies (yellow). Middle panels - Seyfert 2 galaxies (blue) and LINERs (green).  Right panels -  \hii\ galaxies (black) and \hii/AGN composites (light blue). Top panels - intrinsic 2-10 keV luminosity. Middle panels - 12 \mic\ luminosity. Bottom panel - \oiii\ (corrected) luminosity.}
 \label{fig_lumdist}
\end{minipage}
\end{figure*}

Following the optical emission line activity classification we have conducted for our sub-sample, we go on to investigate the luminosity characteristics of the different optical types. Table \ref{table_lum} presents the mean luminosities of each optical type including the standard deviation of the distribution. Plotted in Fig. \ref{fig_lumdist} are histograms of the luminosity distributions for pure \hii\ galaxies and \hii/AGN composites, Seyfert 1s and intermediate Seyfert 1.2-1.9s and Seyfert 2s and LINERs at 2-10 keV (intrinsic), 12 \mic\ and the \oiii\ line. We exclude 3C273 in our luminosity analysis due to its blazar nature. 

In X-rays, it would appear that \hii/AGN composite galaxies, thought to harbour at least a low level AGN, have very similar luminosity distributions to pure \hii\ galaxies. Strict Seyfert 1s have a greater average luminosity than the intermediate types and Seyfert 2s and these in turn have a greater average luminosity than LINERs. A Kolmogorov-Smirnov (K-S) test shows a significant difference between Seyfert 1s and Seyfert 2s at the 99.4\% confidence level, but does not show a significant difference for other associated pairs at the greater than 99\% confidence level. The X-ray luminosities are absorption corrected, and thus, the difference seen between the Seyfert 2s and Seyfert 1s must be intrinsic, rather than due to obscuration.

At 12 \mic, \hii\ galaxies and \hii/AGN composites have the same average luminosity. Strict Seyfert 1s however, show a much larger average 12 \mic\ luminosity than the intermediate type Seyferts and Seyfert 2s. Seyfert 2s again have a greater average luminosity than LINERs. A K-S test shows that Seyfert 1s are distinctly different from both Seyfert 1.2-1.9s  and Seyfert 2s at the 99.99\% confidence level.

Finally, for extinction corrected \oiii\ luminosities, we find a greater average luminosity for \hii/AGN composites when compared to pure \hii\ galaxies.  Seyfert 1s have on average almost a two orders of magnitude greater \oiii\ luminosity than the intermediate Seyferts and Seyfert 2s. Again, we find a greater average luminosity for Seyfert 2s when compared to LINERs, though this may be expected as Seyfert 2s and LINERs are split using the \oiii/H$\beta$ line ratio. A K-S test, however, only shows Seyfert 2s and LINERs to be belonging to different populations at the 99\% confidence level. Furthermore, we note that for $L_{\rm [OIII]}>10^{42}$ \ergs\ there are no pure star forming galaxies. Although selecting AGN above this luminosity would yield sources 100 times more luminous than X-ray selection at $10^{42}$ \ergs, it is still useful to know for sources where standard BPT diagnostics are not possible due to H$\alpha$ having been redshifted out of the optical band and/or X-ray data are not available. In this case the \oiii/H$\beta$ ratio would still be available for AGN/LINER separation.

At all three wavelengths, after accounting for reddening/absorption, Seyfert 1s are intrinsically more luminous than Seyfert 2s. This has been found similarly in the {\it Swift/BAT} hard X-ray sample by \citet{winter10} for \oiii\ and 2-10 keV luminosities. 

\begin{figure}
 \includegraphics[width=90mm]{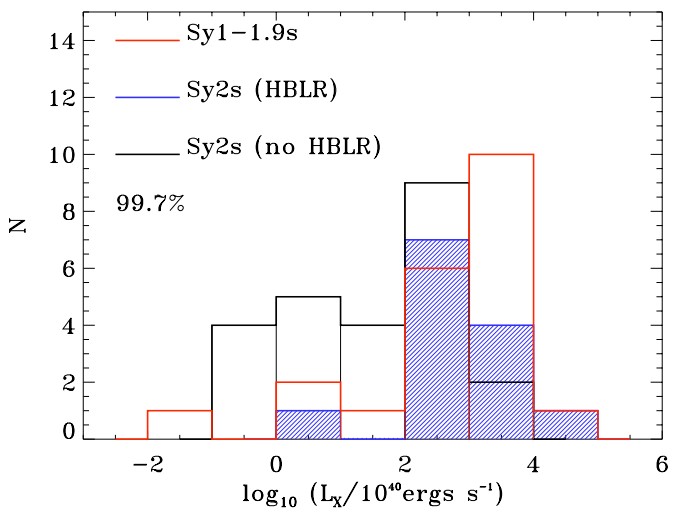}
 \caption{X-ray luminosity distribution of Seyferts 2s with detected HBLRs (blue with hatching), Seyfert 2s with no HBLR detected (black) and Seyfert 1-1.9s (red).}
 \label{fig_bllxdist}
\end{figure}

We also investigate the distribution of luminosities for Seyfert 2s with a detection of an HBLR (either in optical spectropolarimetry, or near-IR spectroscopy) and without a detection.  The mean luminosities of the Seyfert 2s with an HBLR detection is over an order of magnitude higher in X-rays than those without a detection, and a K-S test shows that these distributions are significantly different at the 99.7\% confidence level. Fig. \ref{fig_bllxdist} plots these X-ray luminosity distributions along with the Seyfert 1-1.9 X-ray luminosity distribution for comparison. This results was originally presented by \citet{tran03}, also working on the 12MGS.

\begin{figure*}
\begin{minipage}{180mm}
\includegraphics[width=160mm]{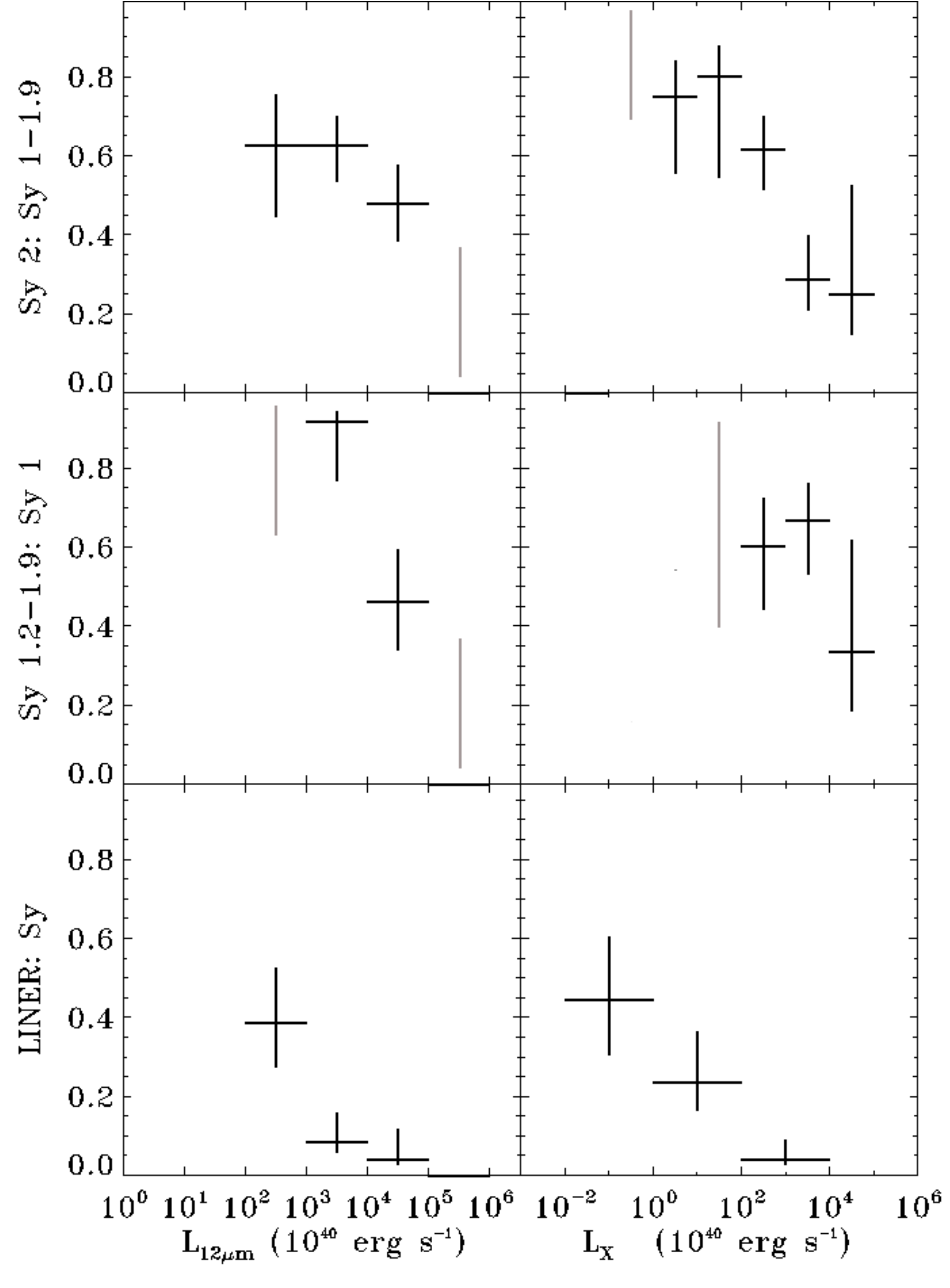}
 \caption{Optical type fractions versus 12 \mic\ luminosity and intrinsic 2-10 keV X-ray luminosity with 68\% confidence intervals plotted as vertical error bars. Lone grey vertical bars represent confidence intervals on the fraction where the computed fraction is 1 or 0 (described in text). Top - Seyfert type 2 fraction of all Seyferts; middle - intermediate Seyfert (type 1.2-1.9) fraction of all type 1 Seyferts; bottom - LINER fraction of all Seyfert and LINERs galaxies. }
 \label{fig_otypefrac}
\end{minipage}
\end{figure*}

Further to the luminosity distributions of these optical types, we also investigate optical type fractions as functions of luminosity. These are presented in Fig. \ref{fig_otypefrac} which shows the Seyfert type 2 fraction of all Seyferts, the Seyfert intermediate type fraction of all type 1 Seyferts and the LINER fraction of all Seyfert and LINER galaxies against 12 \mic\ and intrinsic X-ray luminosity. The vertical error bars we plot here are 68\% confidence intervals calculated using a Bayesian based method presented in \citet{cameron10} which is particularly useful for fractions at or close to 1 or 0. For cases where the fractions are 1 or 0, we plot only the confidence interval, marked by a grey bar. We find here that the Seyfert type 2 fraction is a strong decreasing function of X-ray luminosity, dropping significantly from $\sim$ 60\% to $\sim$ 30\% above 10$^{43}$ \ergs. A weaker and less significant decline is also seen with 12 \mic\ luminosity. We do find however that the fraction of intermediate Seyfert types of all type 1 Seyferts is a strong decreasing function of 12 \mic\ luminosity, though this is not evident in X-ray luminosity. The LINER fraction declines with both 12 \mic\ and X-ray luminosites.

\begin{table}
\centering
\caption{Mean luminosities of the different optical types at 2-10 keV (intrinsic), 12 \mic\ and the \oiii\ line, with their standard deviations, $\sigma$.}
\label{table_lum}
\begin{center}
\begin{tabular}{l c c c c c c}
\hline
Type & $L_{\rm X}$ & $\sigma$ & $L_{\rm 12}$ & $\sigma$ & $L_{\rm OIII}$ & $\sigma$ \\
 & \multicolumn{6}{c}{log$_{10}$ \ergs} \\

\hline
Sy 1			& 43.34 & 0.72 & 44.75 & 0.64 & 43.00 & 1.07 \\
Sy 1.2-1.9		& 42.60 & 1.32 & 43.44 & 0.79 & 41.17 & 1.24 \\
Sy 2			& 41.97 & 1.29 & 43.68 & 0.63 & 41.23 & 1.27 \\
LINERs		& 40.80 & 1.63 & 42.87 & 0.90 & 39.26 & 1.82 \\
\hii			& 40.00 & 0.93 & 43.33 & 0.89 & 39.30 & 1.29 \\
\hii/AGN		& 40.51 & 1.42 & 43.47 & 1.07 & 40.41 & 1.51 \\
\hline
Sy 2 (HBLR) 	& 42.83 & 0.78 & 44.01 & 0.67 & 41.83 & 0.90 \\
Sy 2 (non-HBLR) & 41.51 & 1.27 & 43.50 & 0.54 & 40.91 & 1.34 \\
\hline
\end{tabular}
\end{center}
\end{table}

\section{X-ray properties of the 12MGS}
\label{xrayprop}

We now explore the X-ray properties of our sub-sample according to optical type, building on the analysis in paper I. There are 78/126 (=62\%) galaxies defined as AGN in our sample using our optical definitions, compared to 60/126 (=48\%) using an observed 2-10 keV X-ray luminosity of 10$^{42}$ \ergs\ as a discriminator.  In general we find good agreement between the unambiguous X-ray AGN we found in paper I and the optical definitions we have determined here. All Seyfert 1s, 17/21=81\% of Seyfert 1.2-1.9s and 23/37=62\% of Seyfert 2s have their optical classifications as AGN confirmed in X-rays. For LINERs, 2/10=20\% have $L_{\rm X}>10^{42}$ \ergs\ as do 3/18=17\% composite \hii/AGN galaxies. Finally, none of the 13 \hii\ galaxies present an X-ray luminosity greater than 10$^{42}$ \ergs. 

Table \ref{table_selection} presents the relative number of each optical class that would be solely selected by the X-ray luminosity. Here we also show how adding \nh\ information might aid in determining AGN activity in X-rays, given in columns 4 and 6, as heavily absorbed sources are strong candidates for being AGN. We chose an \nh\ cut off of $\geq 10^{23}$ \cmsq\ as none of the pure \hii\ galaxies in our sample exhibit absorption above this level. We also explore the use of 10$^{41}$ \ergs\ as an effective discriminator for AGN activity in columns 5 and 6. 

In both limiting luminosities, adding the \nh\ information increases the number of AGN selected (from 56 to 64 for $L_{\rm X}>10^{42}$ \ergs\ and from 64 to 69 for $L_{\rm X}>10^{41}$ \ergs), and does not add to the number of non-AGN, except for two additional LINERs which probably harbour an AGN. We should also note that two pure \hii\ defined galaxies, NGC 1482 and NGC 1808 display a large amount of absorption in their X-ray spectra (\nh\ $\simeq 9\times10^{22}$ \cmsq). Though this is slightly less than our 10$^{23}$ \cmsq\ criterion above, it is still a strong indication that these galaxies may host an AGN. It is possible though, that these don't host an AGN, and that \hii\ galaxies can be heavily obscured.
%This supports the use of \nh\ information in classifying AGN in the X-rays.

Lowering the X-ray luminosity to 10$^{41}$ \ergs\ includes a further eight optical AGN types, with the inclusion of only two \hii\ galaxies, which represents a 20\% contamination level for the log$_{10}(L_{\rm X}$/\ergs) =41-42 range. This lower luminosity limit also includes a further two Compton thick sources. This shows that a lower X-ray luminosity of 10$^{41}$ \ergs\ is an effective discriminator for AGN activity.

\begin{table}
\centering

\caption{The number of each optical type that would be selected using various X-ray information. `CT sources' are X-ray sources that are Compton thick. Column (1) gives the optical type; Column (2) gives the number in our \xmm\ sub-sample; Column (3) gives the number that would be selected with $L_{\rm X}>10^{42}$ \ergs; Column (4) gives the number that would be selected with $L_{\rm X}>10^{42}$ \ergs\ OR \nh\ $>10^{23}$ \cmsq; Column (5) gives the number that would be selected with $L_{\rm X}>10^{41}$ \ergs;  Column (6) gives the number that would be selected with $L_{\rm X}>10^{41}$ \ergs\ OR \nh\ $>10^{23}$ \cmsq. $L_{\rm X}$ is the observed 2-10 keV luminosity}
\label{table_selection}
\begin{center}
\begin{tabular}{l c c c c c}
\hline
Type 	& \multicolumn{5}{c}{Number} \\
		& this sample & $L_{\rm X}$ & $L_{\rm X}$+\nh & $L_{\rm X}$ & $L_{\rm X}$+\nh \\
(1) & (2) & (3) & (4) & (5) & (6) \\
\hline
Sy 1		& 12 & 12 & 12 & 12 & 12 \\% & 10 & 10 \\
Sy 1.2-1.9 & 21 & 17 & 18 & 18 & 19 \\% & 19 & 19 \\
Sy 2		& 37 & 23 & 29 & 27 & 31 \\% & 23 & 29 \\
Non Sy HBLR & 8 & 4 & 5 & 7 & 7 \\% & 6 & 6 \\
All Sy+HBLR & \bf 78 & \bf 56 & \bf 64 & \bf 64 & \bf 69 \\% &\bf  58 &\bf  64} \\
\hline
LINERs	& 10 & 2 & 4 & 5 & 5 \\% & 5 & 5 \\
Sy 2/LINERs & 2 & 1 & 1 & 2& 2 \\% & 1 & 1 \\
\hii		& 13 & 0 & 0 & 2 & 2 \\% & 0 & 0 \\
\hii/AGN	& 18 & 3 & 3 & 5 & 5 \\% & 1 & 3 \\
\hline
CT sources & 16 & 11 & & 13 & \\% & 11 & \\
\hline
\end{tabular}
\end{center}
\end{table}

\begin{figure*}
\begin{minipage}{180mm}
 \includegraphics[width=180mm]{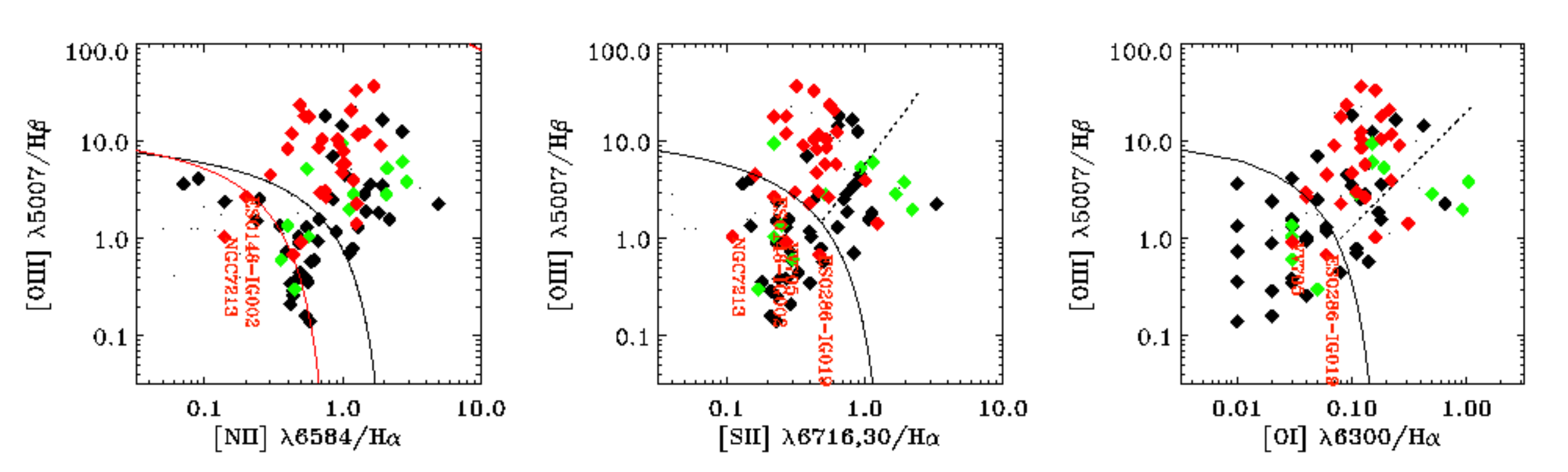}
 \caption{BPT diagrams as in Fig. \ref{fig:mwbanalysis_bpt}, where the colour of each point represents the X-ray luminosity of the galaxy. Red points are unambiguous AGN as defined by their observed X-ray luminosity, with $L_{\rm X} \geq 10^{42}$ \ergs, green points have $10^{41} \leq L_{\rm X} < 10^{42}$ \ergs\  and black points have $L_{\rm X} < 10^{41}$ \ergs. We note the names of galaxies that appear to be AGN from their X-ray luminosity, but have \hii\ line ratios.}\label{fig_xbpt}
\end{minipage}
\end{figure*}

Fig. \ref{fig_xbpt} shows the BPT diagrams of Fig. \ref{fig:mwbanalysis_bpt}, but with X-ray luminosity indicated. Here it can be seen that most of the galaxies with $L_{\rm X} \geq 10^{42}$ \ergs\ lie in the AGN regions of these diagrams. 8/60 (=13\%) galaxies have $L_{\rm X} \geq 10^{42}$ \ergs (observed), but are not unambiguous Seyferts (i.e. Seyfert 1-1.9s or galaxies with a Seyfert 2 classification from all three diagrams). We note VV705, ESO286-IG019, ESO148-IG002, NGC7213 on these diagrams as being X-ray luminous AGN, but with \hii\ line ratios. VV705 and ESO286-IG019 are classed as \hii/AGN composite galaxies using diagram 1, which supersedes the \hii\ classification in the other two diagrams. ESO148-IG002 and NGC7213 are classified as \hii\ on diagrams 1 and 2, but as Seyfert 2 and LINER respectively on diagram 3. We call ESO148-IG002 a \hii/AGN composite and NGC7213 an \hii/LINER. Furthermore, NGC 4388 and MCG 03-58-007 are Sy2/LINERs from their optical line ratios, NGC 6240 is a pure LINER and NGC 6552 has no line ratios available for classifications, but all of these sources are unambiguously powered by an AGN due to their X-ray luminosity. Overall, the X-ray luminosity is in agreement with the optical classification.

\subsection{Continuum}

\begin{figure*}
\begin{minipage}{180mm}
 \includegraphics[width=180mm]{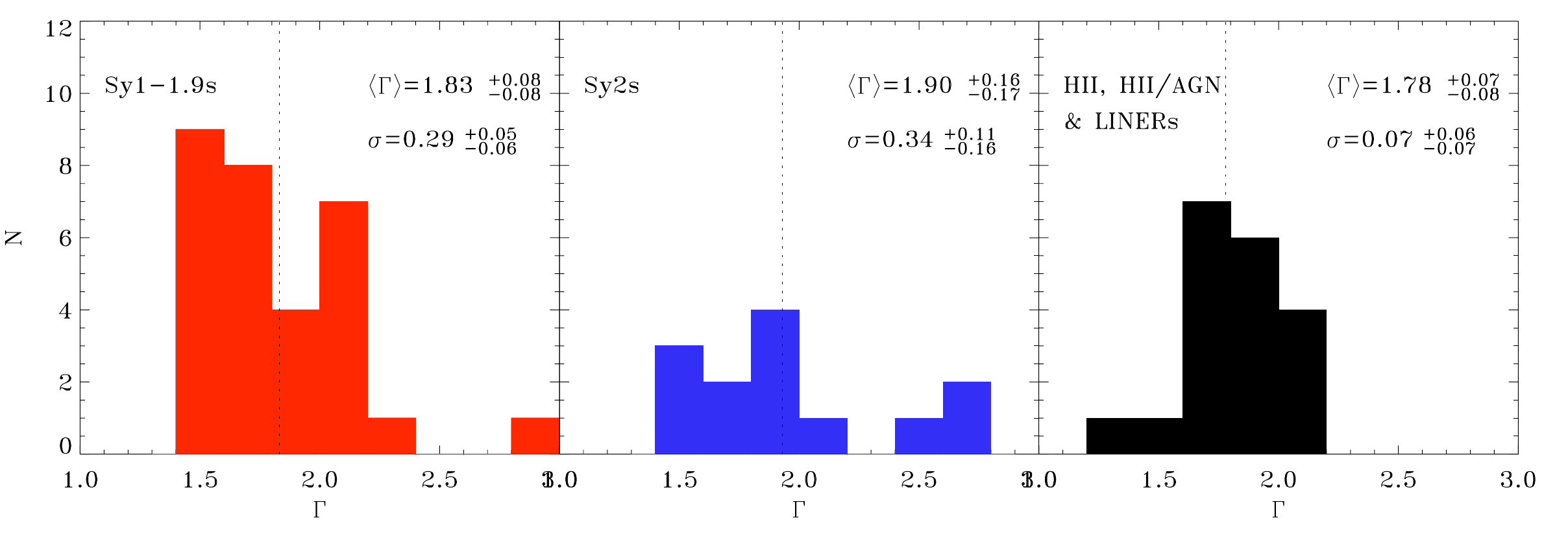}
 \caption{$\Gamma$ distribution of the sources by optical type. Left panel - Seyfert 1-1.9 galaxies. Middle panel - Seyfert 2 galaxies. Right panel - \hii\ galaxies, \hii/AGN composites and LINERs. The means of the distributions are consistent with each other, however, AGN powered sources show a wider range of indices}
 \label{fig_gammdist}
  \end{minipage}
\end{figure*}

Fig. \ref{fig_gammdist} gives the distribution of the X-ray power-law index, $\Gamma$, for the various optical types. We have grouped \hii\ galaxies, \hii/AGN composite galaxies and LINERs into one non-Seyfert category, Seyfert 1-1.9s into another category and Seyfert 2s into the last category. We use the maximum likelihood method of \citet{maccacaro88} to determine both the mean and the intrinsic dispersion for each of these distributions, accounting for the measurement errors. We find that for non-Seyferts, $<\Gamma>=1.78_{-0.08}^{+0.07}$ and $\sigma = 0.07_{-0.07}^{+0.06}$, for Seyfert 1-1.9s, $<\Gamma>=1.83_{-0.06}^{+0.06}$ and $\sigma = 0.29_{-0.06}^{+0.05}$ and for Seyfert 2s, $<\Gamma>=1.90_{-0.17}^{+0.16}$ and $\sigma = 0.34_{-0.16}^{+0.11}$. Statistically, we find that there is no difference in $\Gamma$ between the different optical AGN types. This is in general support of AGN unification schemes as it suggests that the intrinsic properties of the central engines of Seyfert 1s and 2s are the same.
On the other hand, the spectral properties of non-Seyferts are more uniform, showing no significant intrinsic dispersion in their X-ray spectral indices, in contrast to Seyferts.

\subsection{Obscuration}

\begin{figure*}
\begin{minipage}{180mm}
\includegraphics[width=180mm]{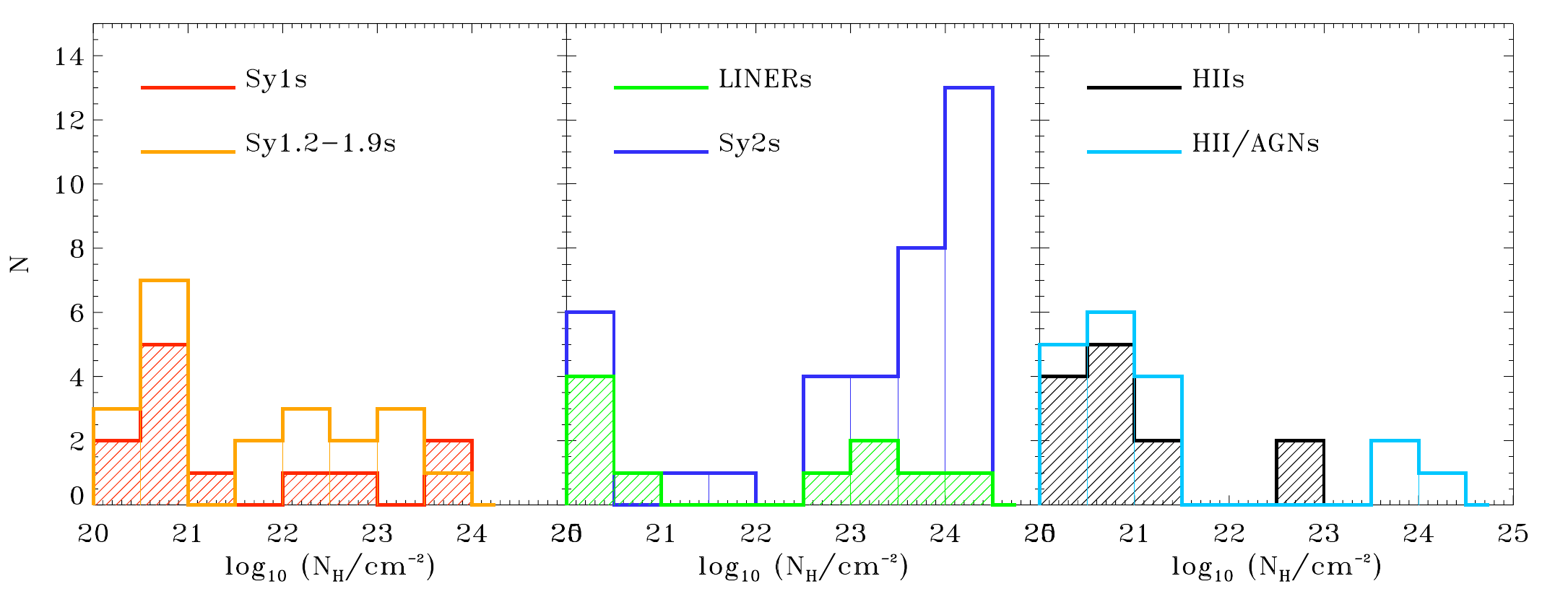}
\caption{\nh\ distribution of the sources by optical type. Left panel - Seyfert 1s (red) and Seyfert 1.2-1.9 galaxies (yellow). Middle panel - Seyfert 2 galaxies (blue) and LINERs (green). Right panel - \hii\ galaxies (black) and \hii/AGN composites (light blue). Seyfert 1s and Seyfert 1.2-1.9s show very similar absorption distributions to each other, while a few of each class show heavy X-ray absorption. Seyfert 2s are mostly heavily absorbed, though a few sources show minimal X-ray absorption. The LINER population has both obscured and unobscured sources. \hii\ and \hii/AGN composites present similar distributions to each other, mostly unobscured, though with some exceptions.}
\label{fig_onhdist}
\end{minipage}
\end{figure*}

Fig. \ref{fig_onhdist} plots the distribution of neutral absorbing columns that were detected in the \xmm\ spectra.  As expected Seyfert 2 galaxies exhibit the largest amounts of absorption in their spectra with an average log$_{10}$ \nh\ = 23.08, whereas Seyfert 1s have an average log$_{10}$ \nh\ = 21.42 and Seyfert 1.2-1.9s have an average log$_{10}$ \nh\ = 21.64. Nonetheless we do note that  a naive interpretation of optical broad line AGN as being X-ray unobscured and narrow lined AGN as being X-ray obscured is incorrect, as 8/37 ($22\pm7\%$) Seyfert 2s show low levels of X-ray absorption \citep[$<10^{22}$ \cmsq, e.g. ][]{pappa01, panessa02} and 4/12 ($33\pm14$\%) Seyfert 1s exhibit high levels of neutral absorption \citep[$>10^{22}$ \cmsq, e.g. ][]{wilkes02}. The uncertainties on the \nh\ values are also consistent with the mismatch. Together these amount to 12/49 ($24\pm6\%$) cases where there is a mismatch between the measured X-ray absorption and the obscuration inferred from the optical type. The absorbed Seyfert 1s tend to be optically reddened, for example IRASF13349+2438 \citep{wills92}, suggesting that the spectrum is attenuated by a dusty torus, as we see in X-rays.  We previously investigated the nature of the unabsorbed Seyfert 2s of the 12MGS in \citet{brightman08}, finding that in about half of cases, it is possible that X-ray absorption has been missed by wide-beam X-ray telescopes due to host galaxy contamination \citep[see also][]{ghosh07,shu10}. The remaining unabsorbed Seyfert 2s are possible `true Seyfert 2s' which are missing the BLR altogether \citep{bianchi08}. Thus in reality, 24\% is an upper limit to the X-ray - optical mismatch. Also we show no difference between the average absorption occurring in strict Seyfert 1 and intermediate Seyferts, as would be expected from models that describe the weakness of the broad lines in the intermediate types due to absorption. 

Furthermore, LINERs show moderate absorption levels on average as log$_{10}$ \nh\ = 21.92, but the distribution is bimodal, with roughly half showing low absorption and half showing heavy absorption. As expected, \hii\ and \hii/AGN composites exhibit low absorption levels with log$_{10}$ \nh\ = 20.99 and 21.24 respectively. However, a few show heavy absorption, so it is a key question as to whether these are in fact AGN or not.

\begin{figure}
\begin{center}
\includegraphics[width=90mm]{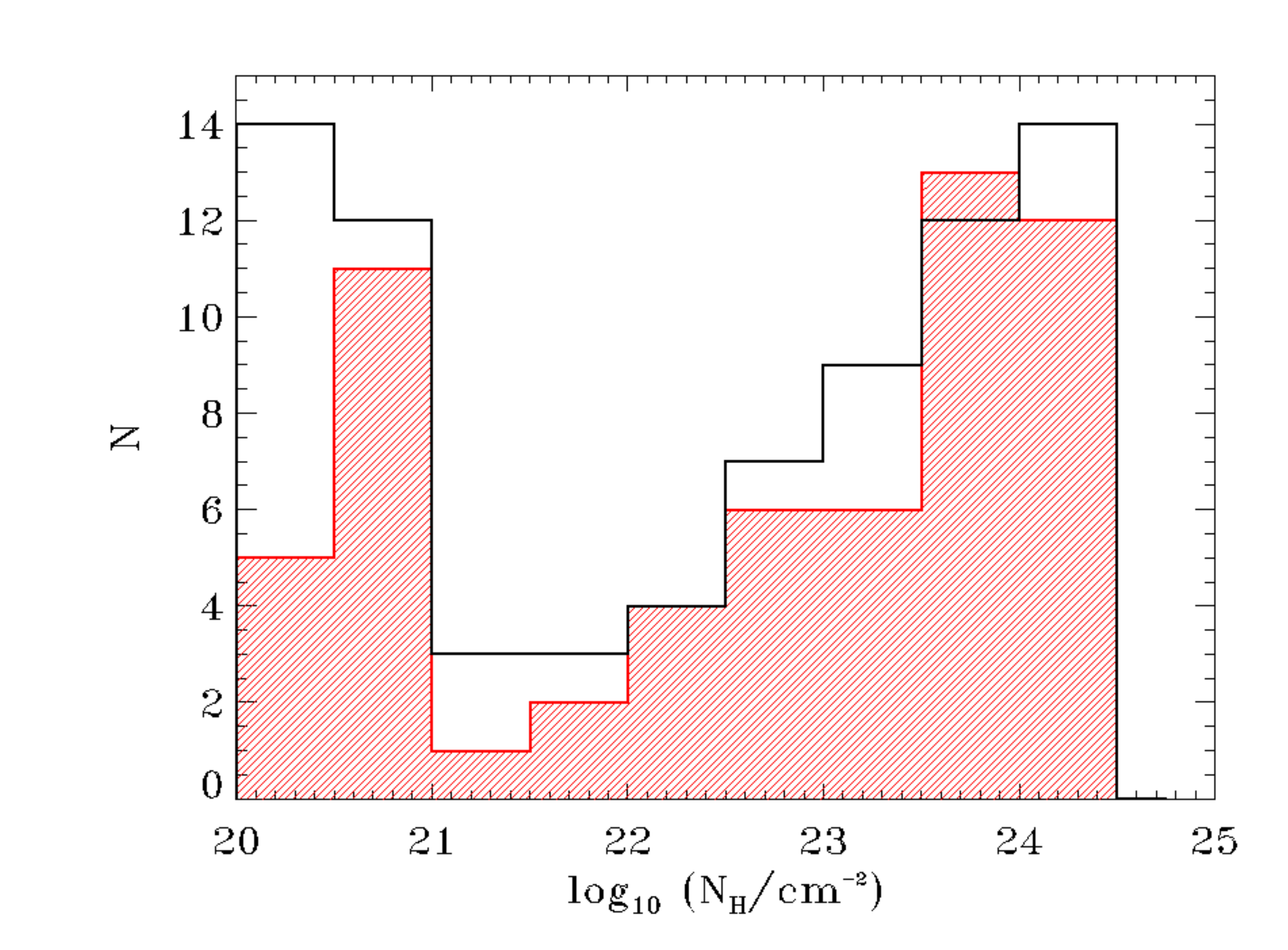}
 \caption{\nh\ distribution for optically defined AGN (black) and X-ray luminous ($L_{\rm X} >10^{42}$ \ergs) AGN (red). }
 \label{fig_oxnhdist}
 \end{center}
% \label{fig:xrb}
\end{figure}

In Fig. \ref{fig_onhdist} we plot the absorption characteristics of the different optical types in our sample. In paper I, we also investigated absorption, but for X-ray luminous ($L_{\rm X} >10^{42}$ \ergs) AGN. Shown in Fig. \ref{fig_oxnhdist} is a comparison plot of the distribution of neutral column densities for X-ray luminous AGN and optically defined AGN. The distributions are fairly similar, with the most notable difference being that optical selection includes a greater proportion of X-ray unabsorbed sources (32/78 = 41\%) over X-ray luminous sources (19/60 = 32\%). In paper I we also showed that the obscured fraction for X-ray sources decreases for luminosities less than 10$^{42}$ \ergs. It follows then that if one selects sources only above this luminosity there is a bias against lower absorbed sources. 

\begin{figure*}
\begin{minipage}{180mm}
\includegraphics[width=160mm]{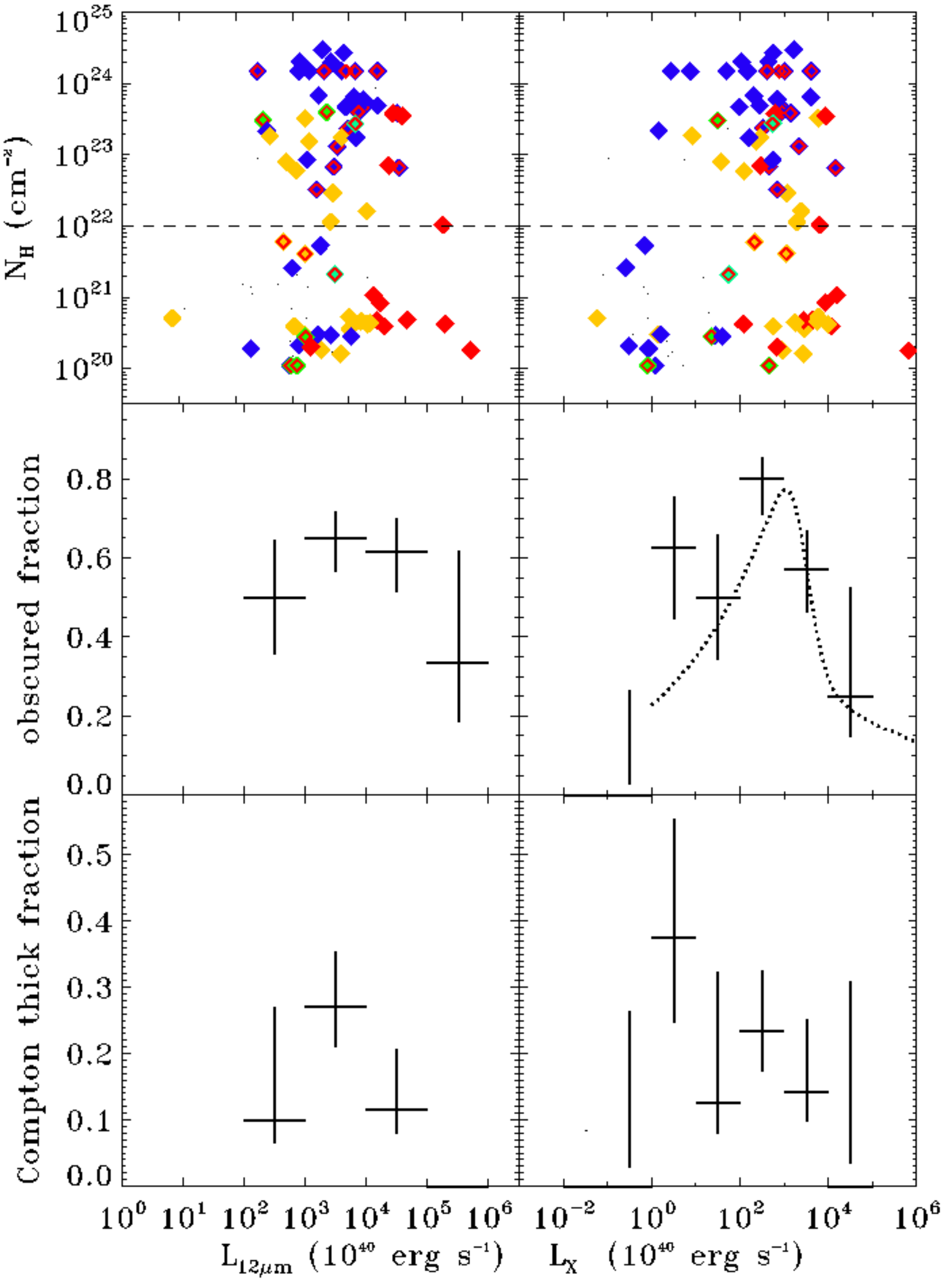}
 \caption{\nh, obscured fraction and Compton thick fraction versus 12\mic\ and intrinsic 2-10 keV luminosity, for AGN types only (non-AGN plotted as dots). Red, yellow and blue diamonds are Seyfert 1s, 1.2-1.9s and 2s respectively, whereas green diamonds are LINERs with an HBLR. Red outlines indicate an HBLR detected in the source. Obscured and Compton thick fractions are calculated in equal logarithmic luminosity bins, as the number of sources with \nh\ $> 10^{22}$ \cmsq\ divided by the total number of sources in that bin, where more than one source exists per bin. The dotted line in the middle right panel is that from Burlon, et al, from their X-ray luminosity function analysis of {\it Swift/BAT} data.} \label{fig_obsfrac}
\end{minipage}
\end{figure*}

We can investigate directly the variation of the obscured fraction for optically defined AGN with both intrinsic X-ray and 12 \mic\ luminosity. Fig. \ref{fig_obsfrac} presents the scatter of \nh\ against $L_{\rm 12\umu m}$ and $L_{\rm X}$, the obscured fraction, defined as the ratio of the number of AGN with \nh\ $>10^{22}$ \cmsq\ to the total number of AGN for each luminosity bin and the Compton thick fraction. We choose to bin in equal steps of luminosity in logarithmic space. The errors on this plot represent 68\% confidence intervals and have been calculated using a Bayesian method particularly useful for fractions at or close to 0 or 1 (described earlier in the text) The lone vertical error bars represent 68 \% confidence intervals on fractions that have been calculated to be 0 or 1. We also plot the obscured fraction against X-ray luminosity as presented by \citet{burlon10} from their X-ray luminosity function analysis of {\it Swift/BAT} data for comparison. As the ratio of 15-55 keV fluxes to the 2-10 keV flux is approximately equal to 1 for a model power-law with $\Gamma=1.9$, we do not scale the X-ray luminosity. The hard X-ray samples extend to higher X-ray luminosity than the 12 \mic\ sample, but the 12 \mic\ sample probes to lower X-ray luminosity.  Our results agree very well here, despite different wavelength selections, and support not only a decrease in the obscured fraction at high luminosities, but also at low luminosities. The Compton thick fraction shows a similar behaviour in X-ray luminosity, though the statistics are not as good. The variation of the obscured fraction with 12 \mic\ luminosity also hints at a decrease at higher luminosities and possibly at lower luminosities, however, the simplest case of non-varying obscured fraction fits the data too. We do see a hint that the Compton thick fraction is dependent on 12 \mic\ luminosity though, suggesting that obscuration is in fact also dependent on 12 \mic\ luminosity.

From our analysis of the \xmm\ spectra of these galaxies, we found a total of 16 Compton thick sources in our sample. All but three of these are optically defined as Seyfert 2, two are \hii/AGN composites (NGC 3690 and ESO 148-IG 002), and the final source, NGC 6552, has not been optically classified here. NGC 6552 is classified as a Seyfert 2 however, by \citet{degrijp92}, though they do not present line ratio data. The Compton thick nature of NGC 6552 was serendipitously discovered by \citet{reynolds94} and \citet{fukazawa94} with an {\it ASCA} observation which shows a reflection dominated spectrum and it also has a HBLR revealed in optical spectropolarimetry \citep{tran01}. Thus, all of our Compton thick sources are AGN. Combining optical and X-ray selection, we find a total of 82 AGN in this sample, which gives us 16/82 (20 $\pm$ 4\%) Compton thick AGN in this sample with 51/82 (62 $\pm$ 5\%) obscured AGN (\nh\ $\geq 10^{22}$ \cmsq).

\begin{figure}
\begin{center}
\includegraphics[width=90mm]{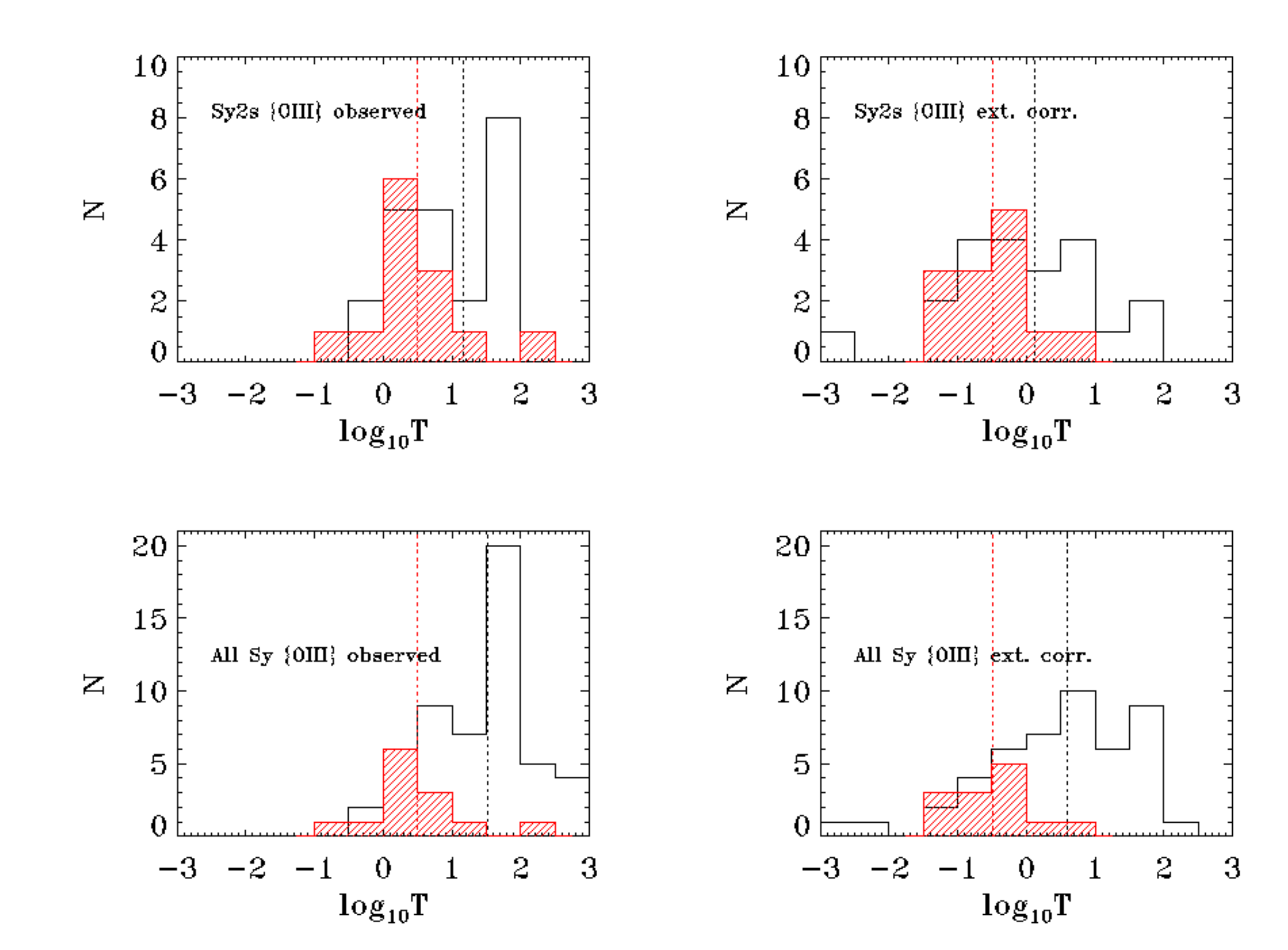}
 \caption{Histograms showing the distribution of the `T' ratio, defined as $F_{\rm X}/F_{\rm [OIII]}$ for Compton thin (black) and Compton thick (red) sources. The top panels show only Seyfert 2, whereas the bottom panels show all Seyferts. The left panels use the observed \oiii\ flux, whereas the right panels use the extinction corrected \oiii\ flux. The X-ray flux is observed in all cases. The dotted lines show the means of the distributions, given with their standard deviations in Table \ref{table_trat}.}
 \label{fig_tratio}
 \end{center}
% \label{fig:xrb}
\end{figure}

Combining X-ray data with data at other wavelengths has been used previously to assess the Compton thick nature of AGN. \citet{bassani99} use the ratio of the X-ray flux to \oiii\ line flux, known as the `T' ratio, to pick out potentially Compton thick candidates not recognised in their X-ray spectra alone. The \oiii\ line is thought to be an isotropic indicator of AGN strength as it is produced in the NLR. However this line is also subjected to extinction along the line of sight and a reddening correction is often applied which is derived from the Balmer decrement. Here we investigate this ratio and its effectiveness at identifying Compton thick AGN. Fig. \ref{fig_tratio} presents the distribution of this ratio for measured Compton thin  (black) and thick (red) sources. We show this for the observed \oiii\ flux and the extinction corrected \oiii\ flux, for Seyfert 2s alone, and for all Seyferts. We present the means and standard deviations of these distributions in Table \ref{table_trat}. Although it can be seen that Compton thick AGN have on average a lower T ratio than Compton thin AGN, the ratio does not separate Compton thick sources from the Compton thin population completely in any case, and many Compton thin AGN have low T ratios. We investigate the Seyferts which are not directly measured to be Compton thick, but have T ratios consistent with Compton thick obscuration (T$<0.1$ is generally taken to be the criterion for a candidate Compton thick AGN \citep{bassani99}). Four of these sources exist, being IRASF07599+6508, MRK0273, NGC 5775 and UGC 09944. IRASF07599+6508 is a Seyfert 1 ULIRG weak in X-rays and also has a large extinction correction ($\rm H\alpha/H\beta$=33.6). MRK 0273 is a heavily obscured Seyfert 2, but not Compton thick and NGC 5775 is also a Seyfert 2 with no apparent indications of heavy obscuration in its X-ray spectrum. NGC 5775 also has an extremely large extinction correction ($\rm H\alpha/H\beta$=126). Finally UGC 09944 is also a Seyfert 2 with a low measured \nh, however, we do note an excess around 6.4 keV, which we can fit with a gaussian with high, but badly constrained EW (3.02$_{-2.51}^{+4.61}$ keV) and unconstrained energy (E=6.50$_{-0.28}^{+31.4}$). This source is potentially a Compton thick Seyfert 2. However, in 2 out of 4 cases, the extinction correction to the \oiii\ flux is likely to have given the source an unreliably low T. In the other two cases, one source is heavily obscured but not Compton thick, which leaves us with only one true Compton thick candidate. Our clear conclusion is that the T ratio is not a reliable Compton thick indicator, especially when uncertainties regarding extinction corrections are considered. It can however, be used to exclude the possibility of a source being Compton thick, if it has a {\it high} T ratio ($>10$ when using extinction corrected \oiii), as these sources have been shown to be almost exclusively Compton thin. We also note that our Compton thick Seyfert 2s have a range of T values that extend to higher values than \citet{bassani99}, up to 6.3 compared to $\sim1$. This may be due to our higher signal-to-noise data, with which we have managed to directly measure the Compton thickness in four of our sources, whereas \citet{bassani99} mostly infer the Compton thickness from a low T and high Fe K$\alpha$ EW.

\begin{table}
\centering

\caption{Mean `T' ratios with their standard deviations, $\sigma$, for Sy 2s, all Sys, using both the observed \oiii\ flux and the extinction corrected \oiii\ flux.}
\label{table_trat}
\begin{center}
\begin{tabular}{l l l l l}

\hline
Type & log$_{10}$T & $\sigma$ & log$_{10}$T & $\sigma$\\
 & \multicolumn{2}{c}{(\oiii\ obs.)} &  \multicolumn{2}{c}{(\oiii\ ext. corr.)} \\

\hline
Compton thin Sy 2s & 1.17 & 0.93 & 0.12 & 1.31 \\
Compton thick Sy 2s & 0.49 & 0.71 & -0.50 & 0.65 \\
Compton thin Sys & 1.51 & 0.89 & 0.59 & 1.22 \\
Compton thick Sys & 0.49 & 0.71 & -0.50 & 0.65 \\
\hline\
\end{tabular}
\end{center}
\end{table}

\section{X-ray - infrared relationship in galaxies}
\label{xrayirrel}

\begin{figure*}
\begin{center}
\includegraphics[width=150mm]{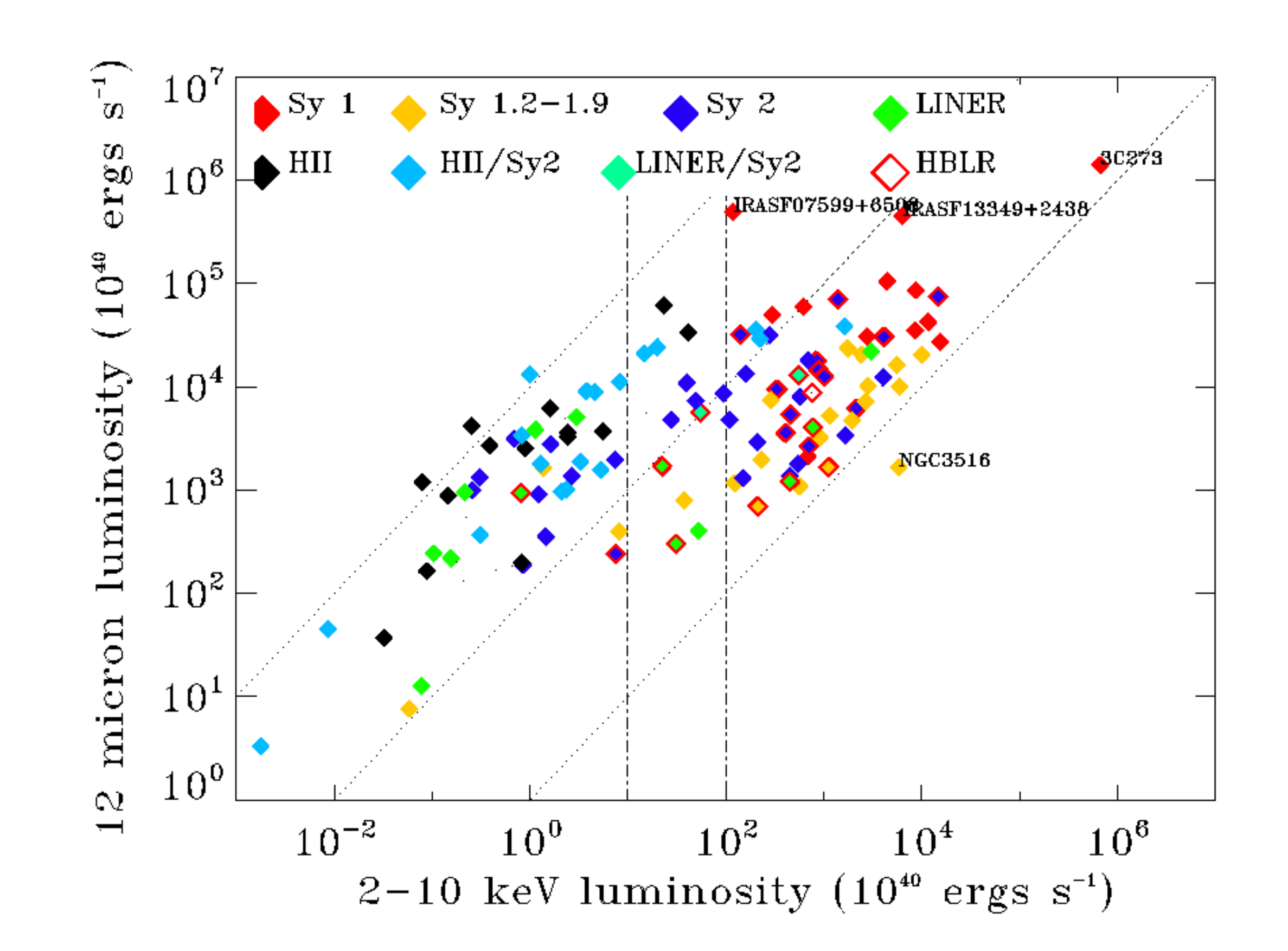}
\includegraphics[width=150mm]{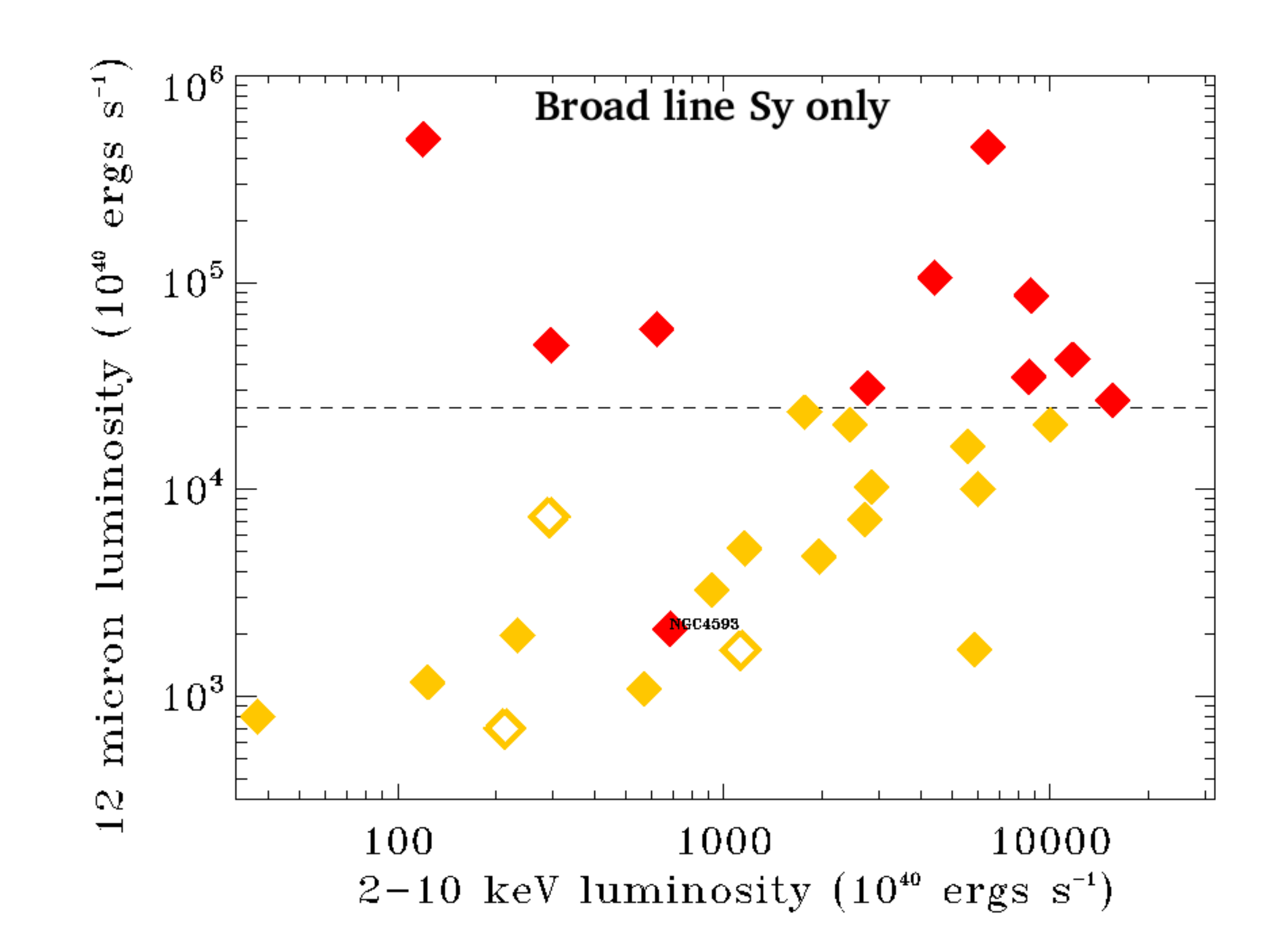}

 \caption{12\mic\ luminosity versus intrinsic 2-10 keV luminosity for our sample. (top) The dotted lines on this diagram represent $L_{\rm 12 \umu m}=L_{\rm X}$, $L_{\rm 12 \umu m}=100\times L_{\rm X}$ and $L_{\rm 12 \umu m}=10^{4}\times L_{\rm X}$. (bottom) Only Seyfert 1-1.9 (broad line Sy) types are plotted here. We show Sy 1.8-1.9s as open yellow symbols.}
 \label{fig_lx12}
 \end{center}
% \label{fig:xrb}
\end{figure*}

\citet{barcons95} and \citet{mckernan09} both find a correlation, though a non-linear one, between X-ray and {\it IRAS} 12 \mic\ luminosities of AGN. These authors attribute some of this to a contribution to the 12 \mic\ luminosity by emission from the host galaxy, but \citet{barcons95} also points to a break down of the unification scheme as this non-linear relationship implies that the covering fraction of the torus is a function of X-ray luminosity. In the top panel of Fig. \ref{fig_lx12} we plot the 12 \mic\ luminosity against the intrinsic X-ray luminosity for all 126 galaxies in our sample, colour coded by optical type. The dotted lines on this diagram represent $L_{\rm 12 \umu m}=L_{\rm X}$, $L_{\rm 12 \umu m}=100\times L_{\rm X}$ and $L_{\rm 12 \umu m}=10^{4}\times L_{\rm X}$. The X-ray luminosities have been corrected for absorption. It is clear here that the relationship between 12 \mic\ luminosity  and X-ray luminosity is non-linear, and there is a large scatter, even for AGN only. Star-forming galaxies also have a much larger  $L_{\rm 12 \umu m}/L_{\rm X}$ than AGN. The dot-dashed lines on this plot  mark the 10$^{41}$ and 10$^{42}$ \ergs\ X-ray luminosities, again showing that all but two galaxies with $L_{\rm X}>10^{41}$ \ergs\ are AGN.

The lower plot in Fig. \ref{fig_lx12} highlights the luminosities of the different type 1 Seyferts. Most striking, is the separation between Seyfert 1 galaxies and Seyfert 1.2-1.9 galaxies in 12 \mic\ luminosity which we discovered previously. Seyfert 1s almost exclusively have $L_{12 \umu m}  > 3 \times 10^{44}$ ergs s$^{-1}$, whereas Seyfert 1.2-1.9 have luminosities exclusively less than this. %Interestingly, Seyfert 2 galaxies show no such difference.

\begin{figure*}
\begin{center}
\includegraphics[width=150mm]{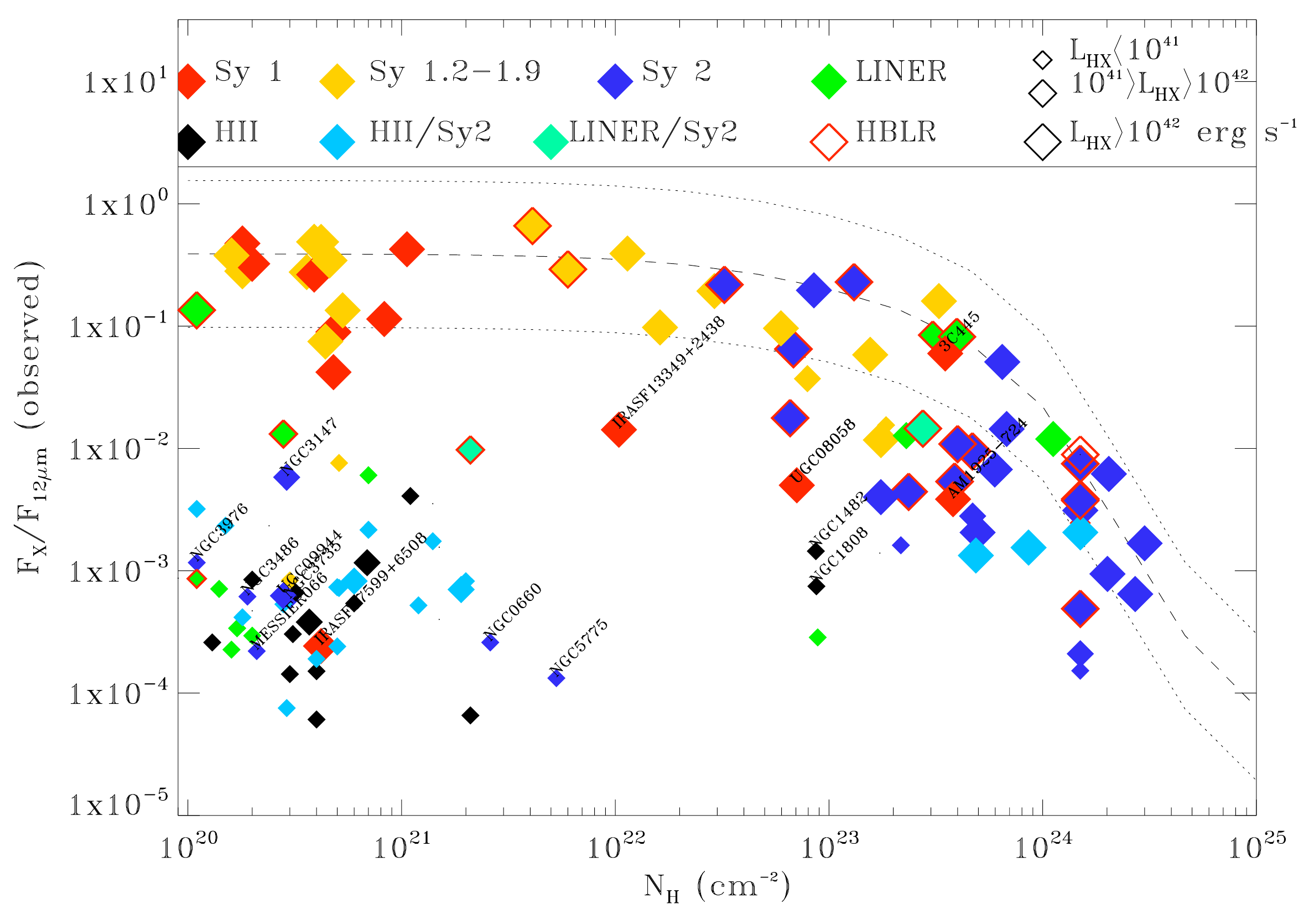}
 \caption{Observed 2-10 keV to 12\mic\ flux ratio versus measured column density for our sample. The colour of each symbol represents the optical type of galaxy, whereas the size of the symbol represents the X-ray luminosity of the galaxy. The dashed line is the ratio of the respective bolometric corrections for AGN for the quantities given, which we find to be $0.40 \pm 0.6$ dex, where the dotted lines indicate the spread in this value. The ratio is also plotted as a function of the X-ray column density, where the 2-10 keV contribution has been suppressed by a factor calculated from our Monte-Carlo model. Named are galaxies where interesting cases where the X-ray obscuration is in contradiction to the optical type (eg. obscured Seyfert 1s)}
 \label{fig_fx12nh}
 \end{center}
% \label{fig:xrb}
\end{figure*}

\citet{horst07} find that Seyfert 1s and Seyfert 2s have the same intrinsic distribution of $L_{\rm MIR}/L_{\rm X}$, which is not expected from smooth dusty torus models, as the 12 \mic\ flux should be suppressed in edge on, Seyfert 2 type systems \citep{pier92}. This is a key result when considering 12\mic\ selection, and using the 12MGS to infer properties of obscuration in local AGN, so we also test if there is any dependence between the X-ray to 12 \mic\ flux ratio and absorption measured in the X-ray band. In Fig. \ref{fig_fx12nh}, we show the observed 2-10 keV flux to the 12 \mic\ flux against the neutral column density measured.  The dashed line we plot  is the ratio of the 12 \mic\ to 2-10 keV bolometric corrections for AGN, $\kappa_{\rm 12 \umu m}/\kappa_{2-10 keV}$.  We use the average $\kappa_{\rm 2-10 keV}$, from \citet{vasudevan07} with the total 1 dex spread that they present, and we calculate $\kappa_{\rm 12\umu m }$, from the bolometric luminosities of the AGN given in \citet{spinoglio95} which we calculate to be 10.1, with a standard deviation of 0.34 dex. This gives $\kappa_{\rm 12 \umu m}/\kappa_{2-10 keV} = 0.40$ with a spread of 0.6 dex represented by the dotted line in each plot. This is consistent with the results of \citet{horst07} and their nuclear 12.3 \mic\ observations (0.42 for Seyfert 1s and 0.36 for Seyfert 2s). The bolometric ratio is also plotted as a function of the X-ray column density, where the 2-10 keV contribution has been suppressed by a factor calculated from our Monte-Carlo model presented in paper I. We place AGN found to be Compton thick in our previous analysis at \nh=$1.5\times10^{24}$ \cmsq, which is a lower limit to their \nh. The symbols are colour coded to represent their optical types.  

We find that most Seyfert 1s have X-ray to MIR flux ratios consistent with the bolometric corrections, as do some Seyfert 2s, however, most Seyfert 2s have a lower than expected ratio, even when taking the suppression of the X-ray flux by absorbing columns into account. Star-forming galaxies are seen to have a much lower ratio than AGN. It is likely that the lower than expected X-ray to MIR flux ratio seen in Seyfert 2s is due to MIR dust emission from their host galaxies which have been heated by star formation.

\section{discussion}\label{discussion}

\subsection{X-ray properties of Seyfert galaxies}

We have presented the average intrinsic X-ray ($L_{\rm X}$ and $\Gamma$) and X-ray absorption (\nh) properties of our sample for each optical class that we have determined here, be it Seyfert 1s, Seyfert 1.2-1.9s, Seyfert 2s, LINERs, \hii\ galaxies or composite \hii/AGN galaxies. For Seyfert galaxies, much of our existing knowledge of these properties comes from analysis of the X-ray selected sample of \citet{piccinotti82}. Here we can compare the properties we derive from our MIR selected sample to those earlier results.
The Seyfert 1s in our sample have an average intrinsic X-ray luminosity of log$_{10}L_{\rm X}=43.3\pm0.72$, $<\Gamma>=1.83_{-0.06}^{+0.06}$ and $\sigma_{\Gamma} = 0.29_{-0.06}^{+0.05}$ ($<\Gamma>$ for Sy 1-1.9s) and log$_{10}$ \nh\ = 21.42. For the X-ray selected sample of Seyfert 1s measured by {\it ASCA} from \citet{nandra97_2}, they find an average log$_{10}L_{\rm X}=43.3\pm0.91$ and $<\Gamma>=1.90\pm0.17$ and $\sigma_{\Gamma} = 0.15\pm0.05$. Our mean luminosities are the same, and our $<\Gamma>$ are consistent with each other, however, we find a significantly broader distribution of $\Gamma$ than in the \citet{nandra97_2} sample.
The Seyfert 2s in our sample have an average intrinsic X-ray luminosity of log$_{10}L_{\rm X}=42.0\pm1.29$, $<\Gamma>=1.90_{-0.17}^{+0.16}$ and $\sigma_{\Gamma} = 0.34_{-0.16}^{+0.11}$ and log$_{10}$ \nh\ = 23.08. {\it ASCA} results from \citet{turner97} for Seyfert 2s, derived from the same X-ray selected sample as \citet{nandra97_2} give $<\Gamma>=2.09\pm0.02$. However, due to the complexities present in fitting the X-ray spectra, of Seyfert 2 galaxies, such as complex absorption, it is not straightforward to compare our results with theirs.

For LINERs, the most comprehensive X-ray spectral study so far has been done by \citet{gmartin09} based on a multi-wavelength selection. The LINERs in our sample have an average intrinsic X-ray luminosity of log$_{10}L_{\rm X}=40.8\pm1.63$ and an average log$_{10}$ \nh\ = 21.92.  \citet{gmartin09}  find that the average 2-10 keV luminosity of their \xmm\ sample is log$_{10}L_{\rm X}=40.3\pm1.5$ and log$_{10}$ \nh\ = 21.93. These parameters are very similar between the two works, despite differing selection techniques, which may suggest homogeneity among the LINER class.

For the \hii\ (\hii/AGN composite) galaxies, we find the average log$_{10}L_{\rm X}=40.0\pm0.93$ ($40.5\pm1.42$) and log$_{10}$ \nh\ = 21.0 (21.2). For the analysis of $\Gamma$ we group together LINERs, \hii\ galaxies, and \hii/AGN composites  and find that $<\Gamma>=1.78_{-0.08}^{+0.07}$ and $\sigma_{\Gamma} = 0.07_{-0.07}^{+0.06}$ for these galaxies. \citet{ptak99} presented an X-ray spectral analysis of a sample of low-luminosity AGN, which consisted mostly of LINERs and starbursts, observed with {\it ASCA}. From their power-law spectral fits they find $<\Gamma>=1.71\pm0.41$, which is consistent with our results from a similar mix of galaxies.

\subsection{AGN selection and classification}

We have found that our X-ray data generally support MIR AGN selection, as we have found a higher fraction of obscured AGN (62 $\pm$ 5\%) than optically selected samples  \citep[$\sim$ 50\%, eg,][]{cappi06} and hard X-ray selected \citep[50\%,][]{tueller08} AGN samples, as well as a higher Compton fraction (discussed below).%Our X-ray also data generally support the classification of AGN using optical emission line diagnostics. However, we do find that a higher fraction of AGN are found using these optical methods (62\% of this sample) than using X-ray luminosity alone (48\% {\bf with $L_{\rm X}>10^{42}$ \ergs), which is of course a very crude method and misses all low luminosity AGN.} Although the 12MGS is independently selected outside of these wavebands, this sample consists of only those galaxies with an \xmm\ observation. We have shown though that this selection has not biased the sub-sample with respect to the parent 12MGS, as the various optical type proportions are generally consistent with each other. 

As for finding AGN in non-Seyfert galaxies, we find an X-ray luminosity in excess of 10$^{42}$ \ergs\ in 17\% of the composite \hii/AGN defined galaxies, whereas we find no such indications in any of the pure \hii\ defined galaxies. The \hii/AGN composites also have on average a greater \oiii\ luminosity than \hii\ galaxies. As the \oiii\ line is used in the classification of these galaxies, this may be interpreted straightforwardly as a selection effect. However, it is mostly the [N {\sc ii}] line which is used to distinguish these from pure \hii\ defined galaxies, so a greater \oiii\ luminosity indicates AGN activity in the \hii/AGN composite galaxies. From infrared SED fitting, \citet{patel11} find 3\% of composite galaxies in the SWIRE, XMMLSS and Lockman Hole surveys require an AGN component, suggesting that X-ray observations are more efficient at identifying AGN activity. Our findings support such a `composite' classification based on optical line raitos.

We also find evidence for AGN activity in a large number (40\%) of LINERs from the X-ray data. This includes NGC 6240, well known for its LINER classification in the optical, but unambiguously powered by a heavily obscured AGN in the X-rays, as {\it Beppo-SAX} showed with the detection of a hard component \citep{mitsuda95}. However \citet{gmartin09} have recently reported that a much larger fraction, 80\%, show indications for AGN activity in LINERs. However, as well as X-ray luminosity and absorption, they also use X-ray morphology, and data at other wavelengths to suggest AGN activity. By contrast, \citet{goulding09} find only a 25\% AGN fraction in IR bright LINERs from the detection of the [NeV] line. It would seem that LINERs are mostly low luminosity AGN, but that in a few cases, they are more powerful Compton thick AGN. The decline in the LINER fraction of Seyfert and LINER galaxies with 12 \mic\ and X-ray luminosity also supports this conclusion.

%We find that at an X-ray luminosity greater than 10$^{42}$ \ergs, 56/78 (=72\%) of optical AGN are selected, with no contamination from star-forming galaxies. If the level of absorption is known, combining this with a \nh\ = 10$^{23}$ \cmsq\ cut off increases this further to 64/78 (=82\%). We note however, NGC 1482 and NGC 1808, both \hii\ defined galaxies, but with large amounts of absorption measured in their X-ray spectra from the detection of a hidden hard transmission component. \citet{jbailon05} do find evidence however, for a low-luminosity AGN from a {\it Chandra} observation of NGC 1808, which disentangles a hard nuclear component from a extra-nuclear star forming region. NGC 1482 is a superwind galaxy, though previous works have failed to attribute this to an AGN \citep{strickland04}. Overall, though, heavy absorption is still highly indicative of AGN activity.

%Furthermore, lowering the luminosity cut off for AGN selection to 10$^{41}$ \ergs\ , and including \nh, the number of optical AGN selected becomes 69/78 (=88\%), with only a very small (3\%) contamination rate from star-forming galaxies. This is particularly useful for deep field X-ray surveys, where 10$^{42}$ \ergs\ is often used to select AGN. As we have shown, this could be lowered to 10$^{41}$ \ergs, with minimal contamination from star-forming galaxies. Being MIR selected, this sample should contain the most luminous star-forming galaxies, which would support our conclusion regarding this selection technique. 

\subsection{AGN obscuration}

The most commonly compared study of the distribution of X-ray absorbing columns in AGN is that by \cite*{risaliti99}, who showed that $\sim75\%$ of their optically selected Seyfert 2s were heavily absorbed (\nh\ $>10^{23}$ \cmsq) and $\sim50\%$ were Compton thick. The authors applied an \oiii\ flux cut off of $4\times10^{-13}$ erg s$^{-1}$ cm$^{-2}$ to their sample however, leaving them with a sample of bright Seyfert 2s, so a direct comparison to their work is not straightforward. However, for the \oiii\ fluxes we have compiled, if we apply the same cut off and include Seyfert 1.8 and 1.9s, we find that 75 $\pm$ 8\% of them are heavily absorbed and 39 $\pm$ 9\% are Compton thick. This agrees well with the \citet{risaliti99} study, though with a smaller, but not significantly so, proportion of heavily obscured Seyfert 2s being Compton thick. Furthermore, some of the Seyfert 2s in the \citet{risaliti99} study have since been shown to be probably not Compton thick \citep{treister09}, so the true Compton thick fraction for these bright Seyferts is probably closer to $\sim40\%$ as our results suggest.

%For all AGN, we find 62 $\pm$ 5\% in our sample to be obscured (\nh\ $>10^{22}$ \cmsq). This is higher than optically selected \citep[$\sim$ 50\%, eg,][]{cappi06} and hard X-ray selected \citep[50\%,][]{tueller08} AGN samples, though only at a 2-$\sigma$ significance. Of the obscured AGN, 16 are Compton thick, four of which have a direct measurement of the \nh. We infer the rest of these from a combination of the reflection fractions and the Fe K$\alpha$ EWs. 

We find a Compton thick fraction of 20 $\pm$ 4\% in the 12 \mic\ sample. \citet{akylas09} find 3/38 (8\%) of their optically selected sample of Seyferts, which has a similar range in $L_{\rm X}$ to our sample ($10^{38}$ to $10^{43}$ \ergs), to be Compton thick from direct measurements of the \nh. However they also note a low T ratio in many other AGN in their sample, which, if they include these as inferred Compton thick objects, would increase their Compton thick fraction to 20\%, in rough agreement with our statistic. Their 8\% statistic is still within 3-$\sigma$ of our own though and can thus be considered as consistent with our results. \citet{cappi06} also find a 20\% Compton thick fraction from their analysis of the same survey. The hard X-ray selected samples also reveal a significantly lower ($>$3-$\sigma$) Compton thick fraction \citep[$\sim$ 4\%, eg,][]{beckmann09}. In paper I we showed the effect of heavy obscuration on X-ray tranmission above 10 keV. At a column density of $\sim4\times10^{24}$ \cmsq\ the observed flux is half that of the intrinsic flux in the 10-40 keV band, and at $\sim10^{25}$ \cmsq\ it is 10\% of that level, presenting a real bias against heavily obscured systems in hard X-ray surveys, and indeed \citet{malizia09} show that when considering only the closest {\it INTEGRAL} sources, the Compton thick fraction they derive is in better agreement. For the {\it Swift/BAT} survey, \citet{burlon10} account for the selection bias in the hard X-ray band, and recover a 20\% Compton thick AGN fraction in doing so. 

%Furthermore 12 \mic\ selection is probably less biased towards heavily obscured objects than optical selection aswell. In order for the optically selected sample of \citet{akylas09} to reach the same Compton thick fraction that we find, they must include all low T ratio AGN as being Compton thick. 

We have shown however, that using the T ratio to consider an AGN as Compton thick is misleading in many cases, especially when taking the extinction correction into account. We found that 2/4 Seyferts with a T ratio consistent with Compton thick obscuration have extremely large Balmer decrements leading to huge extinction corrections, and probably unreliable T ratios. \citet{lamassa10} investigated the reddening correction of \oiii\ for the 12MGS finding that applying the correction based on the Balmer decrement widened the dispersion in the flux ratio of the \oiii\ line to other isotropic indicators, whereas the dispersion tightened for their optically selected sample. They conclude that the extinction correction, which is based on heterogeneous literature values for the 12MGS, overestimate the reddening due to dust.  One Seyfert however, UGC09944, does seem to be a reliable Compton thick candidate based on the T ratio, shown from a large EW but not well constrained Fe K$\alpha$ line. We do find though that a source with a high T ratio can be excluded from being Compton thick as almost all sources which present a high T ratio are Compton thin.

 \subsection{AGN unification}
 
 Our analysis of the $\Gamma$ distribution of Seyfert 1-1.9 and Seyfert 2s shows that there is not a significant difference in the spectral slopes of the two classes, suggesting that the power generation mechanism in the two classes is essentially the same, and thus in support of the unification of the Seyfert types. Several authors have reported a difference in the two distributions, finding that Seyfert 2s tend to have a harder spectral slope \citep[eg,][]{tueller08,middleton08}. These studies have the advantage of data above 10 keV and therefore are less subjected to absorption effects. However, \citet{malizia03} also make use of data above 10 keV, and although they find a systematic difference between the two types initially, this difference disappears when more complex absorption models are used to model the Seyfert 2 data. \citet{dadina08} also draws the same conclusion from {\it Beppo-SAX} data.%They argue that fitting time averaged spectra, where the absorption is not constant throughout the observation, leads to an anomalous estimate of $\Gamma$ in Seyfert 2s, and therefore are able to reconcile the central engines of both Seyfert types in accordance with the unification scheme. \citet{dadina08} also draws the same conclusion from {\it Beppo-SAX} data. For their set of \xmm\ spectra, \citet{cappi06} also find that their distributions of $\Gamma$ are likely to come from the same population for Seyfert 1s and Seyfert 2s, though they find this average to be $\Gamma\sim1.6$, whereas we find $\Gamma\sim1.9$. They do not however use reflection components in their spectral fits, which is the likely difference between our results.

Arguing against unification schemes are the finding of cases where Seyfert 1s are X-ray absorbed and Seyfert 2s are X-ray unabsorbed, and finding that Seyfert 1s are intrinsically more powerful than Seyfert 2s at 12 \mic, \oiii\ and X-ray wavelengths. Furthermore, we also show that AGN obscuration depends on X-ray luminosity, finding that rather than declining with $L_{\rm X}$, the obscured fraction peaks at $L_{\rm X}\sim10^{42}$ \ergs. Other authors have previously reported that the obscured fraction of AGN decreases at high luminosities \citep[eg.][]{ueda03, hasinger08}, but their samples do not generally extend into the lower luminosity range probed by the 12MGS. The decline in the obscured fraction at high luminosities is supported by the decline we see in the Seyfert 2 fraction above $\sim10^{43}$ \ergs. This has also been previously reported \citep[e.g. ][in \oiii\ luminosity]{simpson05}, though this is still a subject of debate \citep[see ][ for a discussion on this]{lawrence10}.  As for the decline in the obscured fraction towards low luminosities, \citet{akylas09} also find evidence for less obscuration at low luminosities, as do \citet{zhang09}, both for nearby AGN. \citet{burlon10} similarly find a peak in the obscured fraction for the {\it Swift/BAT} selected AGN from luminosity function analysis. The variation of the obscured fraction with 12 \mic\ luminosity is not as statistically significant as seen in X-rays, and is consistent with a non-varying obscured fraction, but better statistics are needed to test this definitively. The Compton thick fraction does, however, seem to vary with 12 \mic\ luminosity, suggesting that obscuration does in fact depend on 12 \mic\ luminosity as well as X-ray luminosity. 

%We find that Seyfert 1s are intrinsically more powerful than Seyfert 2s at 12 \mic, \oiii\ and X-ray wavelengths, which contradicts unified schemes which base the sole differences between the two types on the orientation of the observer. Furthermore, we also find that the detection of hidden broad lines in Seyfert 2s is dependent on X-ray luminosity, with HBLR Sy2s having on average a higher X-ray luminosity than non-HBLR Sy2s. This supports the findings of \citet{tran03}, who show that HBLR Seyfert 2s are intrinsically more powerful than non-HBLR Seyfert 2s, including in X-rays. 

\citet{hopkins09} attribute the disappearance of the BLR to low accretion rate, radiatively inefficient systems as observationally found by \citet{nicastro03} who show that the detection rate of HBLRs in Seyfert 2s decreases with accretion rate. Modelling of outflows in low luminosity AGN by \citet{elitzur06} predicts that both the torus and the BLRs disappear at low bolometric luminosities, which would explain both the decline we see in the obscured fraction at low X-ray luminosities and the lower incidence of HBLRs at low X-ray luminosities. 

Another interesting difference our data reveal is that strict Seyfert 1s have a distinctly different 12 \mic\ luminosity distribution from the intermediate type 1.2-1.9s Seyferts, which is shown to be highly statistically significant  from a K-S test. This is also seen in the decline of the intermediate type fraction with 12 \mic\ luminosity. The progression from strict type 1 through to type 1.5 is dependent on the ratio of the total H$\beta$ flux to the \oiii\ flux, i.e. the relative strengths of the BLR and NLR, where Seyfert 1.5s have the weakest BLR to NLR ratio. The difference in the relative strengths between these Seyfert sub-types may be due to reddening of the BLR, or due to an intrinsic difference in the ionising flux \citep{tran92}. A difference in 12 \mic\ luminosity between the Seyfert 1s and intermediate Seyferts suggests that the 12 \mic\ luminosity is in some way connected to the relative strength of the BLR to the NLR in unobscured AGN. It should be noted that the X-ray luminosity distributions of Seyfert 1s and intermediate Seyferts are not shown to be significantly different. If we then presume that the X-ray luminosity is an indicator of nuclear power; higher 12 \mic\ luminosities in Seyfert 1s might suggest a higher covering fraction of the torus in Seyfert 1s with respect to intermediate Seyferts. This in turn suggests that the strength of the BLR is dependent on the torus covering fraction, and implies an origin in the torus for the BLR. This supports the model presented by \citet{gaskell09}, which describes the BLR as the inner part of the torus, where material has lost angular momentum and the dust has sublimated. This subsequently forms into an accretion disk and is accreted by the black hole.

In conclusion, it seems that our data suggest some intrinsic link between the torus, the BLRs and the accretion disk, and finds good cause for modifications to be made to the simplest AGN unification schemes.

\subsection{Implications for the X-ray background}

The predictions of the recent X-ray background synthesis models of \citet{gilli07} claim to match the observed Compton thick fraction as found by hard X-ray telescopes. However, since publication of the hard X-ray samples, the Compton thick fraction has fallen due to ongoing observations \citep{beckmann09}. \citet{treister09} argue that the discrepancy between observations and models is due to degeneracies in the models used to fit the background spectrum, in particular, the normalisation of the Compton reflection component. They argue that only direct observations can be used to constrain the properties of Compton thick AGN in the local universe. Alternatively, or in addition, consideration of blazars in the synthesis models lead to a lower required CT AGN fraction \citep{draper09}. As we have shown, the 12MGS is well suited to directly constraining the Compton thick fraction of AGN in the local universe due to its representative nature and the relative lack of bias towards heavily obscured systems, so we conclude that the true Compton thick fraction of local AGN is only $\sim20\%$. 

Furthermore, the intrinsic dispersion on $\Gamma$ is an important parameter when considering XRB models. \citet{gilli07} directly investigated the effect of introducing a non zero $\sigma_{\Gamma}$ into the synthesis models, and evaluated the effect of varying the size of the dispersion. They found that the larger the dispersion, the greater contribution to the 30 keV peak from unobscured AGN there was. For their baseline model they use  $\sigma_{\Gamma}=0.2$ based on results from \citet{mateos05} on faint sources in the Lockman Hole. Here we find that for local AGN, $\sigma_{\Gamma}\simeq0.3$, which would introduce a greater contribution to the 30 keV peak from unobscured AGN than the Gilli et al models produce, and hence require a smaller contribution from Compton thick AGN.

\subsection{Implications for high redshift studies}

Mid-IR observations by {\it Spitzer} are being used extensively for deep field, high redshift studies of AGN. This is due to the fact that much of the reprocessed primary energy of the AGN is re-emitted in the mid-IR.  Our local study at 12 \mic\ corresponds exactly to the {\it Spitzer/MIPS} 24 \mic\ selected sources at z=1, and provides valuable insight into the nature of these sources. Building on results from \citet{horst07} and finding that 12 \mic\ selected AGN contain a higher fraction of Compton thick sources than optically or hard X-ray selected samples, we have concluded that mid-IR selection is indeed ideally suited to selecting AGN and is relatively unbiased against obscuration, bolstering these deep field efforts. We have found that using a lower X-ray luminosity of 10$^{41}$ \ergs\ rather than 10$^{42}$ \ergs\ will increase the number of AGN selected in X-ray surveys, with minimal inclusion of star forming galaxies, which is valuable information for the high-z searches. 

Additionally, the 24 \mic\ band of Spitzer/MIPS is often used to select `infrared excess' (IRX) sources, where the flux in the 24 \mic\ band is in excess compared to a wavelength which is more subjected to absorption, such as the optical R band or X-rays \citep[eg. ][]{fiore08, georgantopoulos09}. These IRX sources are candidates for high-z Compton thick AGN, but as we have shown for our local sources, there are cases where low X-ray to mid-IR AGN (infrared excess sources) are not in fact heavily obscured and have intrinsically low ratios (eg. IRASF 07599+6508, NGC 3147, IRASF 01475-0740.). Recently \citet{georgakakis10} investigated the nature of z$\sim$2 IRX sources and showed that only a small fraction of them displayed tentative evidence for being Compton thick.

\section{Conclusions}
\label{conclusions}

In summary, we have uniformly determined the optical types for the 126 galaxies that form a sub-sample of the 12MGS, for which we have good X-ray data from paper I using BPT line ratio diagnostics. We have conducted a study comparing the optical and X-ray selection methods and have also characterised the optical types in the X-ray and we have investigated X-ray properties of the sample by optical type.

The main conclusions from this study have been that:
\begin{itemize}

\item strict Seyfert 1s are distinctly more powerful at X-ray luminosities than Seyfert 2s and distinctly more powerful at 12 \mic\ luminosities than both intermediate Seyferts and Seyfert 2s.

\item Seyfert 2 galaxies with a detection of an HBLR show a significantly higher X-ray luminosity than those without a detection, supporting the findings of \citet{tran03}.

\item The Seyfert 2 fraction of Seyfert galaxies is a strong decreasing function of X-ray luminosity dropping from $\sim$60\% to $\sim$30\% at 10$^{43}$ \ergs.

\item our X-ray data are in general support of the classification of AGN using optical emission lines, though we find seven `hidden' AGN from X-ray data in galaxies optically defined as \hii/AGN composite or LINER, and two further candidates in the \hii\ defined NGC 1482 and NGC 1808 due to their large detected absorption columns.

\item \hii/AGN composites show a higher average \oiii\ luminosity than pure \hii\ galaxies indicating AGN power in the composites, plus 17\% show X-ray indications of AGN power, supporting claims that these galaxies harbour at least a low level active nucleus.

\item X-ray indications of AGN power, being heavy absorption and/or high X-ray luminosity, are found in 40\% of MIR selected LINERs.

\item using a lower X-ray luminosity of 10$^{41}$ \ergs, rather than the widely used 10$^{42}$ \ergs, to select optical AGN is effective, with only a 3\% contamination rate by star-forming galaxies. Including heavily obscured X-ray sources with \nh$>10^{23}$ \cmsq\ also adds to the number of AGN selected with the exclusion of star-forming galaxies.

\item we find general support for AGN unification schemes due to the distribution of power-law indices, $\Gamma$, for Seyfert 1s and 2s being consistent with each other implying that the power generation mechanism is the same in both. We also find that on average the \nh\ measured in Seyfert 2s is higher than Seyfert 1s, as expected from unification schemes.

\item however, in 24\% of cases the absorption measured in the X-ray spectra does not correspond directly with that implied in the optical band from the visibility of the BLRs.

\item a luminosity dependent modification to the AGN unified scheme is required. This is to account for the decrease in the obscured fraction at high X-ray luminosities ($L_{\rm X} > 10^{43}$ \ergs), possibly due to the recession of the torus at these high source powers; and also a decrease in both the obscured fraction and HBLR detection at low X-ray luminosities ($L_{\rm X} < 10^{42}$ \ergs).

\item 12 micron selected galaxies contain a higher fraction of obscured (62 $\pm$ 5\%) and Compton thick (20 $\pm$ 4\%) AGN than hard X-ray and optically selected samples, supporting MIR selection as a relatively unbiased method of selecting AGN.

%\item we find rough agreement in the \nh\ distribution of Seyfert 2s with that of \cite*{risaliti99} when we apply the same \oiii\ flux cut off that they do, though we advocate a lower Compton thick fraction than they report.

%\item we also find a lower Compton thick fraction than X-ray background models imply, though it is believed that the XRB models over predict the Compton thick fraction, mostly due to model degeneracies. 

\item use of the `T' ratio to find candidate Compton thick sources can often be unreliable, partly due to large extinction corrections. We can support the use of the ratio to exclude Compton thickness though, as a high ratio almost exclusively belongs to Compton thin sources. 

\item our work on the locally selected 12MGS is important for high redshift AGN studies, especially for IR selected samples, as we have shown that this selection method is relatively unbiased, and that X-ray luminosity and accurate \nh\ information can be valuable for selecting AGN.
\end{itemize}

\section{Acknowledgements}

\bibliographystyle{mn2e}
\bibliography{bibdesk}

\setcounter{table}{0}
\onecolumn
%\begin{sidewaystable}
\begin{landscape}
\centering
%\begin{scriptsize}
%\begin{tabular}{l l l l l l l l l l r@{}l}\\
\begin{longtable}{l r r r r r r r r l l l l}\\
\caption{Optical line ratio data compiled from the literature and nuclear activity classification based on the scheme of \citet{kewley06}. Also included are the observed fluxes of the \oiii\ emission line, the Balmer decrement, and the derived absorption corrected \oiii\ fluxes. Column (1) Galaxy name; Column (2) the emission line flux ratio \oiii/H$\beta$; Column (3) the emission line flux ratio [O {\sc i}]/H$\alpha$; Column (4) the emission line flux ratio [N {\sc ii}]/H$\alpha$; Column (5) the emission line flux ratio [S {\sc ii}]/H$\alpha$; Column (6) the observed flux of the \oiii\ $\lambda5007$ line in $10^{-16}$ \ergs; Column (7)  the Balmer decrement H$\alpha$/H$\beta$; Column (8) the flux of the \oiii\ line corrected for absorption using the Balmer decrement; Column (9) reference for the line ratio data: 1=\citet{armus89}, 2=\citet{baan98}, 3=\citet{corbett03}, 4=\citet{degrijp92}, 5=\citet{goncalves99}, 6=\citet{ho97}, 7=\citet{kewley01}, 8=\citet{kim95}, 9=\citet{kong02}, 10=\citet{kopylov74}, 11=\citet{maia87}, 12=Rodriguez, et al. In prep, 13=\citet{sekiguchi93}, 14=\citet{vaceli97}, 15=\citet{veroncetty86}, 16=SDSS; Column (10) the activity classification based on diagram 1;  Column (11) the activity classification based on diagram 2; Column (12) the activity classification based on diagram 3; Column (13) the adopted activity classification.}
\label{table_optdat}
\\
\hline
\\
Name & \oiii/H$\beta$ & [O {\sc i}]/H$\alpha$ &  [N {\sc ii}]/H$\alpha$ &  [S {\sc ii}]/H$\alpha$ & F$_{[OIII]}$ & H$\alpha$/H$\beta$ & F$_{[OIII]}$ & ref & \multicolumn{4}{c}{classification} \\
 & & & & & (obs) & & (cor) & & 1 & 2 & 3 & adopted \\
\\
(1) & (2) & (3) & (4) & (5)  & (6)  & (7) & (8)  & (9)  & (10)  & (11) & (12) & (13) \\

\hline
MRK 0335 	&	0.25	&	-	&	-	&	-	&	2250	&	-	&	-	&	12	&	-	&	-	&	-	&	Sy1.2	\\
NGC 0017 	&	2.86	&	0.11	&	1.20	&	0.52	&	91	&	20.6	&	23800	&	12	&	AGN	&	Sy2	&	Sy2	&	Sy2	\\
NGC 0150 	&	1.32	&	0.03	&	0.55	&	0.27	&	-	&	-	&	-	&	7	&	H{\sc ii}/AGN	&	H{\sc ii}	&	H{\sc ii}	&	H{\sc ii}/AGN	\\
NGC 0214 	&	-	&	-	&	-	&	-	&	-	&	-	&	-	&		&		&		&		&		\\
NGC 0262 	&	9.17	&	0.26	&	0.95	&	-	&	5170	&	3.2	&	5470	&	12	&	AGN	&	-	&	Sy2	&	Sy2	\\
UGC 00545 	&	0.24	&	-	&	0.15	&	-	&	743	&	2.4	&	743	&	12	&	H{\sc ii}	&	-	&	-	&	Sy1	\\
NGC 0424 	&	4.55	&	0.06	&	0.30	&	0.16	&	7000	&	2.9	&	7000	&	14	&	AGN	&	H{\sc ii}	&	Sy2	&	Sy2	\\
NGC 0526A 	&	11.52	&	0.25	&	0.71	&	0.56	&	2470	&	3.0	&	2470	&	12	&	AGN	&	Sy2	&	Sy2	&	Sy1.5	\\
NGC 0513 	&	3.05	&	0.11	&	0.75	&	0.46	&	490	&	3.1	&	490	&	12	&	AGN	&	Sy2	&	Sy2	&	Sy2	\\
NGC 0520 	&	0.74	&	0.01	&	0.39	&	0.29	&	41	&	4.2	&	126	&	6	&	H{\sc ii}	&	H{\sc ii}	&	H{\sc ii}	&	H{\sc ii}	\\
NGC 0660 	&	2.53	&	0.05	&	0.85	&	0.43	&	84	&	14.3	&	7520	&	6	&	AGN	&	Sy2	&	Sy2	&	Sy2	\\
2MASX J01500266- 	&	5.25	&	-	&	0.55	&	-	&	535	&	7.5	&	7190	&	12	&	AGN	&	-	&	-	&	Sy2/LINER	\\
NGC 0695 	&	0.30	&	0.05	&	0.45	&	0.17	&	-	&	7.7	&	-	&	8	&	H{\sc ii}	&	H{\sc ii}	&	H{\sc ii}	&	H{\sc ii}	\\
NGC 1052 	&	2.00	&	0.92	&	1.12	&	2.22	&	2110	&	2.9	&	2110	&	12	&	AGN	&	LINER	&	LINER	&	LINER	\\
MESSIER 077 	&	9.13	&	0.07	&	1.88	&	0.36	&	163000	&	4.3	&	438000	&	12	&	AGN	&	Sy2	&	Sy2	&	Sy2	\\
ARP 118 	&	12.57	&	0.12	&	1.45	&	0.63	&	465	&	6.5	&	4160	&	12	&	AGN	&	Sy2	&	Sy2	&	Sy2	\\
MCG -02-08-039 	&	18.14	&	0.18	&	0.53	&	0.27	&	1730	&	6.9	&	18100	&	12	&	AGN	&	Sy2	&	Sy2	&	Sy2	\\
NGC 1194 	&	23.77	&	0.09	&	0.49	&	0.56	&	229	&	16.5	&	31400	&	16	&	AGN	&	Sy2	&	Sy2	&	Sy2	\\
NGC 1291 	&	-	&	-	&	-	&	-	&	219	&	-	&	-	&		&		&		&		&		\\
NGC 1313 	&	-	&	-	&	-	&	-	&	-	&	-	&	-	&		&		&		&		&		\\
NGC 1316 	&	-	&	-	&	-	&	-	&	210	&	-	&	-	&		&		&		&		&		\\
NGC 1320 	&	10.49	&	0.12	&	0.71	&	0.43	&	1240	&	5.6	&	6860	&	12	&	AGN	&	Sy2	&	Sy2	&	Sy2	\\
NGC 1365 	&	1.83	&	0.04	&	0.48	&	0.16	&	620	&	8.8	&	13100	&	15	&	H{\sc ii}/AGN	&	H{\sc ii}	&	H{\sc ii}	&	Sy1.8	\\
NGC 1386 	&	16.67	&	0.24	&	1.94	&	0.82	&	7800	&	2.8	&	7800	&	14	&	AGN	&	Sy2	&	Sy2	&	Sy2	\\
NGC 1482 	&	0.21	&	-	&	0.42	&	0.29	&	-	&	-	&	-	&	7	&	H{\sc ii}	&	H{\sc ii}	&	-	&	H{\sc ii}	\\
3C 120 	&	1.26	&	0.01	&	0.05	&	-	&	3820	&	6.9	&	39600	&	12	&	H{\sc ii}	&	-	&	H{\sc ii}	&	Sy1	\\
NGC 1614 	&	1.05	&	0.03	&	0.57	&	0.22	&	912	&	11.1	&	38900	&	12	&	H{\sc ii}/AGN	&	H{\sc ii}	&	H{\sc ii}	&	H{\sc ii}/AGN	\\
MRK 0618 	&	19.50	&	0.07	&	0.71	&	0.41	&	1410	&	-	&	-	&	7,12	&	AGN	&	Sy2	&	Sy2	&	Sy1	\\
NGC 1672 	&	0.39	&	-	&	0.47	&	0.24	&	998	&	6.8	&	12600	&	12	&	H{\sc ii}	&	H{\sc ii}	&	-	&	H{\sc ii}	\\
NGC 1667 	&	5.82	&	0.13	&	0.99	&	0.52	&	603	&	10.3	&	20600	&	12	&	AGN	&	Sy2	&	Sy2	&	Sy2	\\
NGC 1808 	&	0.16	&	0.02	&	0.54	&	0.21	&	145	&	14.2	&	16200	&	7,12	&	H{\sc ii}	&	H{\sc ii}	&	H{\sc ii}	&	H{\sc ii}	\\
ESO 362- G 018 	&	1.79	&	0.05	&	0.13	&	0.04	&	3650	&	4.0	&	7660	&	12	&	H{\sc ii}	&	H{\sc ii}	&	H{\sc ii}	&	Sy1.5	\\
2MASX J05210136- 	&	36.85	&	0.12	&	1.69	&	0.32	&	799	&	14.5	&	74200	&	12	&	AGN	&	Sy2	&	Sy2	&	Sy2	\\
2MASX J05580206- 	&	0.52	&	-	&	-	&	-	&	701	&	9.1	&	16400	&	12	&	-	&	-	&	-	&	Sy1	\\
IC 0450 	&	2.12	&	-	&	0.10	&	-	&	7000	&	7.6	&	97000	&	4	&	H{\sc ii}	&	-	&	-	&	Sy1.5	\\
UGC 03973 	&	1.23	&	0.02	&	0.12	&	0.11	&	14	&	6.4	&	120	&	16	&	H{\sc ii}	&	H{\sc ii}	&	H{\sc ii}	&	Sy1.2	\\
IRASF07599+6508	&	0.30	&	-	&	-	&	-	&	33	&	33.6	&	36500	&	12	&	-	&	-	&	-	&	Sy1	\\
NGC 2639 	&	3.37	&	0.39	&	3.62	&	1.77	&	234	&	4.3	&	596	&	12	&	AGN	&	LINER	&	LINER	&	Sy1.9	\\
NGC 2655 	&	3.83	&	1.04	&	2.91	&	1.93	&	391	&	5.0	&	1590	&	6	&	AGN	&	LINER	&	LINER	&	LINER	\\
IC 2431 	&	0.61	&	0.03	&	0.36	&	0.30	&	62	&	4.7	&	263	&	16	&	H{\sc ii}	&	H{\sc ii}	&	H{\sc ii}	&	H{\sc ii}	\\
MRK 0704 	&	0.49	&	-	&	-	&	-	&	1410	&	7.9	&	22200	&	12	&	-	&	-	&	-	&	Sy1.5	\\
NGC 2841 	&	1.85	&	0.17	&	1.84	&	1.13	&	109	&	3.3	&	130	&	12	&	AGN	&	LINER	&	LINER	&	LINER	\\
UGC 05101 	&	2.29	&	0.08	&	1.27	&	0.40	&	45	&	19.7	&	10300	&	12	&	AGN	&	Sy2	&	Sy2	&	Sy2	\\
NGC 2992 	&	10.85	&	0.16	&	0.84	&	0.59	&	2930	&	7.3	&	36300	&	12	&	AGN	&	Sy2	&	Sy2	&	Sy1.9	\\
MESSIER 081 	&	1.14	&	0.21	&	0.37	&	0.23	&	2260	&	5.8	&	13900	&	12	&	H{\sc ii}/AGN	&	H{\sc ii}	&	LINER	&	Sy1.8	\\
MESSIER 082 	&	0.36	&	0.01	&	0.56	&	0.18	&	316	&	25.0	&	146000	&	6	&	H{\sc ii}/AGN	&	H{\sc ii}	&	H{\sc ii}	&	H{\sc ii}/AGN	\\
NGC 3079 	&	3.60	&	0.18	&	1.59	&	0.86	&	18	&	23.6	&	7030	&	12	&	AGN	&	LINER	&	Sy2	&	Sy2	\\
NGC 3147 	&	6.14	&	0.15	&	2.71	&	1.14	&	172	&	5.3	&	828	&	12	&	AGN	&	LINER	&	Sy2	&	Sy2	\\
NGC 3227 	&	2.71	&	0.08	&	0.43	&	0.22	&	6180	&	6.3	&	50400	&	12	&	H{\sc ii}/AGN	&	H{\sc ii}	&	Sy2	&	Sy1.5	\\
NGC 3310 	&	0.95	&	0.04	&	0.66	&	0.26	&	340	&	4.8	&	1200	&	6	&	H{\sc ii}/AGN	&	H{\sc ii}	&	H{\sc ii}	&	H{\sc ii}/AGN	\\
NGC 3486 	&	4.52	&	0.09	&	1.04	&	0.93	&	131	&	3.3	&	159	&	12	&	AGN	&	LINER	&	Sy2	&	Sy2	\\
NGC 3516 	&	0.42	&	-	&	0.06	&	-	&	3630	&	3.0	&	3630	&	12	&	H{\sc ii}	&	-	&	-	&	Sy1.5	\\
MESSIER 066 	&	2.89	&	0.13	&	1.44	&	0.74	&	295	&	5.9	&	1960	&	12	&	AGN	&	LINER	&	Sy2	&	Sy2	\\
NGC 3690 	&	1.37	&	0.03	&	0.40	&	0.25	&	492	&	5.9	&	3230	&	6	&	H{\sc ii}/AGN	&	H{\sc ii}	&	H{\sc ii}	&	H{\sc ii}/AGN	\\
NGC 3735 	&	7.06	&	0.05	&	0.85	&	0.38	&	374	&	6.3	&	2940	&	12	&	AGN	&	Sy2	&	Sy2	&	Sy2	\\
NGC 3976 	&	3.50	&	0.10	&	1.95	&	0.83	&	77	&	4.4	&	210	&	12	&	AGN	&	LINER	&	Sy2	&	Sy2	\\
NGC 3982 	&	14.54	&	0.42	&	0.99	&	0.64	&	1810	&	3.5	&	2480	&	12	&	AGN	&	Sy2	&	Sy2	&	Sy2	\\
NGC 4013 	&	0.71	&	0.11	&	1.13	&	0.83	&	7	&	2.0	&	7	&	6	&	AGN	&	LINER	&	LINER	&	LINER	\\
ARP 244 	&	0.26	&	0.04	&	0.44	&	0.23	&	65	&	4.5	&	251	&	15	&	H{\sc ii}	&	H{\sc ii}	&	H{\sc ii}	&	H{\sc ii}	\\
NGC 4051 	&	1.59	&	-	&	0.26	&	0.14	&	3950	&	3.2	&	4490	&	12	&	H{\sc ii}	&	H{\sc ii}	&	-	&	Sy1.5	\\
NGC 4151 	&	2.51	&	-	&	0.16	&	-	&	126000	&	2.8	&	126000	&	12	&	H{\sc ii}	&	-	&	-	&	Sy1.5	\\
NGC 4214 	&	3.67	&	0.01	&	0.07	&	0.13	&	3470	&	2.8	&	3470	&	6	&	H{\sc ii}	&	H{\sc ii}	&	H{\sc ii}	&	H{\sc ii}	\\
NGC 4253 	&	1.68	&	-	&	0.23	&	-	&	4540	&	5.0	&	18900	&	4	&	H{\sc ii}	&	-	&	-	&	Sy1.5	\\
MESSIER 099 	&	0.90	&	0.02	&	0.48	&	0.23	&	29	&	6.3	&	226	&	6	&	H{\sc ii}/AGN	&	H{\sc ii}	&	H{\sc ii}	&	H{\sc ii}/AGN	\\
MESSIER 100 	&	0.79	&	0.11	&	1.18	&	0.48	&	52	&	4.6	&	160	&	6	&	AGN	&	H{\sc ii}	&	LINER	&	LINER	\\
NGC 4388 	&	12.04	&	6.99	&	0.43	&	0.27	&	4890	&	4.8	&	17600	&	12	&	AGN	&	Sy2	&	LINER	&	LINER	\\
NGC 4414 	&	0.58	&	0.14	&	0.59	&	0.50	&	19	&	3.0	&	19	&	6	&	H{\sc ii}/AGN	&	H{\sc ii}	&	LINER	&	H{\sc ii}/LINER	\\
NGC 4449 	&	2.41	&	0.02	&	0.14	&	0.23	&	1780	&	3.2	&	2570	&	6	&	H{\sc ii}	&	H{\sc ii}	&	H{\sc ii}	&	H{\sc ii}	\\
3C 273 	&	-	&	-	&	-	&	-	&	-	&	-	&	-	&	12	&	-	&	-	&	-	&	Sy1	\\
NGC 4490 	&	2.55	&	0.12	&	0.25	&	0.71	&	33	&	6.3	&	262	&	6	&	H{\sc ii}/AGN	&	LINER	&	Sy2	&	H{\sc ii}/AGN	\\
MESSIER 088 	&	5.31	&	0.19	&	2.10	&	0.94	&	369	&	3.6	&	559	&	6	&	AGN	&	Sy2	&	Sy2	&	Sy2	\\
NGC 4559 	&	0.35	&	0.03	&	0.42	&	0.40	&	9	&	3.7	&	19	&	6	&	H{\sc ii}	&	H{\sc ii}	&	H{\sc ii}	&	H{\sc ii}	\\
MESSIER 090 	&	1.18	&	0.06	&	0.90	&	0.40	&	552	&	5.0	&	2250	&	6	&	AGN	&	H{\sc ii}	&	H{\sc ii}	&	H{\sc ii}/AGN	\\
MESSIER 058 	&	2.87	&	0.50	&	2.07	&	1.69	&	724	&	3.4	&	941	&	12	&	AGN	&	LINER	&	LINER	&	LINER	\\
NGC 4593 	&	22.91	&	0.12	&	1.26	&	0.29	&	1630	&	5.8	&	10400	&	7,12	&	AGN	&	Sy2	&	Sy2	&	Sy1	\\
MESSIER 104 	&	1.57	&	0.18	&	2.19	&	1.07	&	22	&	3.3	&	27	&	6	&	AGN	&	LINER	&	LINER	&	LINER	\\
NGC 4631 	&	1.53	&	0.03	&	0.24	&	0.23	&	23	&	3.0	&	28	&	6	&	H{\sc ii}	&	H{\sc ii}	&	H{\sc ii}	&	H{\sc ii}	\\
NGC 4666 	&	1.31	&	0.06	&	1.30	&	0.64	&	-	&	7.7	&	-	&	8	&	AGN	&	LINER	&	H{\sc ii}	&	H{\sc ii}/LINER	\\
NGC 4725 	&	-	&	-	&	-	&	-	&	-	&	-	&	-	&		&		&		&		&		\\
UGC 08058 	&	0.37	&	-	&	-	&	-	&	1650	&	-	&	-	&	12	&	-	&	-	&	-	&	Sy1	\\
NGC 4968 	&	20.88	&	0.21	&	1.15	&	0.60	&	1340	&	9.6	&	37000	&	12	&	AGN	&	Sy2	&	Sy2	&	Sy2	\\
NGC 5005 	&	2.27	&	0.65	&	4.94	&	3.31	&	473	&	2.6	&	473	&	6	&	AGN	&	LINER	&	LINER	&	LINER	\\
MESSIER 063 	&	1.89	&	-	&	1.48	&	0.75	&	44	&	5.6	&	243	&	6	&	AGN	&	LINER	&	-	&	LINER	\\
MCG -03-34-064 	&	11.72	&	0.22	&	1.29	&	0.46	&	15300	&	3.4	&	19600	&	12	&	AGN	&	Sy2	&	Sy2	&	Sy2	\\
NGC 5170 	&	-	&	-	&	-	&	-	&	-	&	-	&	-	&		&		&		&		&		\\
NGC 5194	&	12.64	&	0.15	&	2.72	&	0.89	&	1630	&	8.9	&	36000	&	12	&	AGN	&	Sy2	&	Sy2	&	Sy2	\\
ESO 383- G 035 	&	0.71	&	-	&	0.18	&	-	&	753	&	7.3	&	9150	&	4	&	H{\sc ii}	&	-	&	-	&	Sy1.2	\\
MESSIER 083 	&	0.29	&	0.02	&	0.44	&	0.21	&	336	&	6.1	&	3070	&	15	&	H{\sc ii}	&	H{\sc ii}	&	H{\sc ii}	&	H{\sc ii}	\\
IRASF13349+2438	&	0.40	&	-	&	-	&	-	&	688	&	14.3	&	61900	&	12	&	-	&	-	&	-	&	Sy1	\\
NGC 5256 (S)	&	2.65	&	0.13	&	0.76	&	0.55	&	322	&	-	&	-	&	12	&	AGN	&	Sy2	&	Sy2	&	Sy2	\\
NGC 5253 	&	4.17	&	0.03	&	0.09	&	0.15	&	68000	&	2.9	&	68000	&	14	&	H{\sc ii}	&	H{\sc ii}	&	Sy2	&	H{\sc ii}/AGN	\\
MRK 0273 	&	5.87	&	0.13	&	1.04	&	0.62	&	1800	&	9.2	&	44300	&	12	&	AGN	&	Sy2	&	Sy2	&	Sy2	\\
IC 4329A 	&	0.64	&	0.01	&	-	&	0.02	&	2550	&	11.8	&	129000	&	12	&	-	&	H{\sc ii}	&	H{\sc ii}	&	Sy1.2	\\
UGC 08850 (E)	&	8.30	&	-0.02	&	0.40	&	0.45	&	7570	&	-	&	-	&	12	&	AGN	&	Sy2	&	-	&	Sy2	\\
NGC 5506 	&	8.59	&	0.12	&	0.67	&	0.53	&	4370	&	8.9	&	98400	&	12	&	AGN	&	Sy2	&	Sy2	&	Sy2	\\
NGC 5548 	&	10.09	&	0.36	&	0.88	&	0.66	&	7340	&	1.3	&	7340	&	6	&	AGN	&	Sy2	&	Sy2	&	Sy1.5	\\
NGC 5775 	&	18.41	&	0.10	&	0.75	&	0.65	&	6	&	125.9	&	332000	&	16	&	AGN	&	Sy2	&	Sy2	&	Sy2	\\
2MASX J15115979- 	&	1.27	&	-	&	0.09	&	0.06	&	1910	&	8.6	&	38600	&	12	&	H{\sc ii}	&	H{\sc ii}	&	-	&	Sy1	\\
VV 705 	&	0.92	&	0.03	&	0.50	&	0.27	&	118	&	7.9	&	1850	&	16	&	H{\sc ii}/AGN	&	H{\sc ii}	&	H{\sc ii}	&	H{\sc ii}/AGN	\\
UGC 09944 	&	9.54	&	0.15	&	0.99	&	0.22	&	1460	&	6.1	&	10500	&	12	&	AGN	&	Sy2	&	Sy2	&	Sy2	\\
2MASX J15504152- 	&	17.92	&	0.08	&	0.57	&	0.22	&	1600	&	7.2	&	18600	&	12	&	AGN	&	Sy2	&	Sy2	&	Sy2	\\
NGC 6240 	&	1.43	&	0.31	&	1.26	&	1.24	&	76	&	16.0	&	9540	&	12	&	AGN	&	LINER	&	LINER	&	LINER	\\
NGC 6286 	&	0.45	&	0.08	&	0.49	&	0.33	&	-	&	-	&	-	&	2	&	H{\sc ii}/AGN	&	H{\sc ii}	&	H{\sc ii}	&	H{\sc ii}/AGN	\\
NGC 6552 	&	-	&	-	&	-	&	-	&	-	&	-	&	-	&		&		&		&		&		\\
AM 1925-724 	&	5.25	&	0.14	&	1.15	&	0.72	&	-	&	-	&	-	&	7	&	AGN	&	Sy2	&	Sy2	&	Sy1	\\
NGC 6810 	&	0.60	&	-	&	0.62	&	0.30	&	65	&	12.7	&	4100	&	12	&	H{\sc ii}/AGN	&	H{\sc ii}	&	-	&	H{\sc ii}/AGN	\\
NGC 6890 	&	33.33	&	0.16	&	1.25	&	0.43	&	980	&	13.3	&	71400	&	15	&	AGN	&	Sy2	&	Sy2	&	Sy2	\\
MRK 0509 	&	0.40	&	-	&	0.18	&	-	&	5400	&	4.0	&	11100	&	4	&	H{\sc ii}	&	-	&	-	&	Sy1.2	\\
ESO 286-IG 019 	&	0.68	&	0.06	&	0.44	&	0.47	&	-	&	-	&	-	&	7	&	H{\sc ii}/AGN	&	H{\sc ii}	&	H{\sc ii}	&	H{\sc ii}/AGN	\\
NGC 7090 	&	-	&	-	&	-	&	-	&	-	&	6.7	&	-	&		&		&		&		&		\\
NGC 7172 	&	4.77	&	0.10	&	0.99	&	0.45	&	40	&	3.5	&	57	&	12	&	AGN	&	Sy2	&	Sy2	&	Sy2	\\
NGC 7213 	&	1.04	&	0.16	&	0.14	&	0.11	&	2380	&	7.9	&	37300	&	12	&	H{\sc ii}	&	H{\sc ii}	&	LINER	&	H{\sc ii}/LINER	\\
IC 5169 	&	1.58	&	0.03	&	0.68	&	0.28	&	100	&	-	&	-	&	7,12	&	H{\sc ii}/AGN	&	H{\sc ii}	&	H{\sc ii}	&	H{\sc ii}/AGN	\\
NGC 7252 	&	-	&	-	&	-	&	-	&	-	&	-	&	-	&		&		&		&		&		\\
3C 445 	&	2.89	&	0.02	&	-	&	0.04	&	2520	&	7.7	&	37100	&	12	&	-	&	H{\sc ii}	&	H{\sc ii}	&	Sy1	\\
NGC 7314 	&	1.27	&	0.01	&	0.08	&	0.05	&	479	&	4.8	&	1700	&	12	&	H{\sc ii}	&	H{\sc ii}	&	H{\sc ii}	&	Sy1.9	\\
MCG -03-58-007 	&	7.91	&	-	&	1.02	&	-	&	1270	&	3.8	&	2390	&	12	&	AGN	&	-	&	-	&	Sy2/LINER	\\
NGC 7469 	&	0.44	&	-	&	-	&	-	&	4540	&	-	&	-	&	12	&	-	&	-	&	-	&	Sy1.2	\\
NGC 7479 	&	3.97	&	0.22	&	1.20	&	1.01	&	140	&	4.9	&	533	&	12	&	AGN	&	LINER	&	Sy2	&	Sy2	\\
ESO 148-IG 002 	&	2.69	&	0.04	&	0.20	&	0.22	&	282	&	9.7	&	8090	&	12	&	H{\sc ii}	&	H{\sc ii}	&	Sy2	&	H{\sc ii}/AGN	\\
NGC 7552 	&	0.14	&	0.01	&	0.58	&	0.23	&	-	&	-	&	-	&	7	&	H{\sc ii}	&	H{\sc ii}	&	H{\sc ii}	&	H{\sc ii}	\\
NGC 7582 	&	2.99	&	0.04	&	0.69	&	0.31	&	4530	&	7.0	&	49000	&	12	&	AGN	&	Sy2	&	Sy2	&	Sy2	\\
NGC 7674 	&	10.55	&	0.18	&	0.92	&	0.53	&	4780	&	3.4	&	6210	&	12	&	AGN	&	Sy2	&	Sy2	&	Sy2	\\
NGC 7714 	&	1.35	&	0.01	&	0.35	&	0.15	&	-	&	-	&	-	&	7	&	H{\sc ii}/AGN	&	H{\sc ii}	&	H{\sc ii}	&	H{\sc ii}/AGN	\\
NGC 7771 	&	1.04	&	0.04	&	0.48	&	0.41	&	-	&	-	&	-	&	8	&	H{\sc ii}/AGN	&	H{\sc ii}	&	H{\sc ii}	&	H{\sc ii}/AGN	\\
MRK 0331 	&	0.39	&	0.03	&	0.54	&	0.27	&	84	&	8.3	&	1540	&	12	&	H{\sc ii}/AGN	&	H{\sc ii}	&	H{\sc ii}	&	H{\sc ii}/AGN	\\
\hline
\end{longtable}
\end{landscape}
%\end{sidewaystable}

\begin{landscape}
\centering
\begin{longtable}{l l l r r r r r }\\
\caption{This table presents multi-wavelength data for our sample, which has been in the analysis in this chapter. Column (1) Galaxy name; Column (2) Optical type as determined in Table \ref{table_optdat}; Column (3) Broad line region information taken from \citet{veroncetty06}. h: hidden BLR detected in optical polarised light. i: broad lines detected in the near-IR, b: broad lines detected in the optical spectrum of a LINER or \hii\ region; Column (4) Logarithm of the observed 2-10 keV luminosity (\ergs); Column (5) Logarithm of the 12 \mic\ luminosity (\ergs);  Column (6)  Logarithm of the absorption corrected \oiii\ luminosity (\ergs); Column (7) log$_{10}$ \nh\ measured, \cmsq; Column (8) log$_{10}$(T=F$_{2-10}$/F$_{[OIII]}$) where F$_{2-10}$ is the observed X-ray flux, and F$_{[OIII]}$ is the absorption corrected \oiii\ flux.}
\label{table_tab2}
\\
\hline
\\
Name & Type & BLR & $L_{\rm X}$ & $L_{\rm 12 \umu m}$ & $L_{\rm [OIII]}$ & \nh & T  \\
(1) & (2) & (3) & (4) & (5)  & (6)  & (7) & (8)  \\

\hline
MRK0335&	Sy1.2	&&43.45&44.01&-&20.56&-\\
NGC0017&	Sy2	&&41.98&43.94&42.32&23.67&-0.93\\
NGC0150&	H{\sc ii}/AGN	&&40.37&43.00&-&20.18&-\\
NGC0214&		&&41.19&43.74&-&23.24&-\\
NGC0262&	Sy2	&h&43.34&43.79&41.44&23.12&1.71\\
UGC00545&	Sy1	&&43.65&45.02&41.82&20.68&1.82\\
NGC0424&	Sy2	&h&42.52&43.98&41.34&23.37&0.29\\
NGC0526A&	Sy1.5	&&43.29&43.68&41.31&22.06&1.96\\
NGC0513&	Sy2	&h&42.66&43.73&40.62&22.83&1.92\\
NGC0520&	H{\sc ii}	&&39.95&43.40&39.21&20.49&0.68\\
NGC0660&	Sy2	&&39.41&43.00&40.11&21.42&-0.70\\
2MASXJ01500266&	Sy2/LINER	&h&41.74&43.75&41.71&21.32&0.04\\
NGC0695&	H{\sc ii}	&&41.62&44.52&-&20.84&-\\
NGC1052&	LINER	&h&41.49&42.48&40.07&23.49&1.34\\
MESSIER077&	Sy2	&h&42.15&44.51&42.14&24.18&-0.95\\
ARP118&	Sy2	&&43.61&44.09&41.90&23.81&0.90\\
MCG-02-08-039&	Sy2	&h&42.93&44.25&42.57&23.67&-0.34\\
NGC1194&	Sy2	&&42.32&43.46&42.11&23.83&-0.49\\
NGC1291&		&&39.65&42.23&-&20.20&-\\
NGC1313&		&&39.32&42.13&-&21.18&-\\
NGC1316&		&&40.33&42.97&-&20.38&-\\
NGC1320&	Sy2	&&42.65&43.14&41.08&24.31&-0.16\\
NGC1365&	Sy1.8	&&42.46&43.87&40.94&23.24&1.00\\
NGC1386&	Sy2	&i&40.88&42.38&40.16&24.18&-0.39\\
NGC1482&	H{\sc ii}	&&40.74&43.57&-&22.94&-\\
3C120&	Sy1	&&44.19&44.43&43.00&21.03&1.06\\
NGC1614&	H{\sc ii}/AGN	&&41.17&44.32&42.34&21.28&-1.17\\
MRK0618&	Sy1	&&43.44&44.49&-&20.68&-\\
NGC1672&	H{\sc ii}	&&39.59&43.43&40.73&20.48&-1.15\\
NGC1667&	Sy2	&&42.76&43.91&42.03&24.43&-1.31\\
NGC1808&	H{\sc ii}	&&40.39&43.52&40.59&22.94&-0.20\\
ESO362-G018&	Sy1.5	&&42.37&43.30&41.42&23.19&0.64\\
2MASXJ05210136&	Sy2	&h&44.17&44.88&43.50&22.82&-0.37\\
2MASXJ05580206&	Sy1	&&44.07&44.63&42.64&20.59&1.41\\
IC0450&	Sy1.5	&&43.06&43.72&42.89&22.46&0.11\\
UGC03973&	Sy1.2	&&43.78&44.00&40.13&20.72&3.00\\
IRASF07599+6508&	Sy1	&&42.08&45.69&44.34&20.62&-2.26\\
NGC2639&	Sy1.9	&&40.14&43.22&40.21&20.48&-0.08\\
NGC2655&	LINER	&&41.72&42.61&39.89&23.36&0.82\\
IC2431&	H{\sc ii}	&&41.37&44.79&41.19&20.57&0.18\\
MRK0704&	Sy1.5	&&43.39&44.31&42.64&22.21&0.66\\
NGC2841&	LINER	&&39.19&42.34&38.10&20.15&1.09\\
UGC05101&	Sy2	&&42.45&44.50&42.57&23.70&-0.76\\
NGC2992&	Sy1.9	&i&43.05&43.23&41.68&21.61&1.37\\
MESSIER081&	Sy1.8	&&38.77&40.89&37.86&20.71&0.90\\
MESSIER082&	H{\sc ii}/AGN	&&40.11&43.25&40.20&20.71&-0.08\\
NGC3079&	Sy2	&&40.87&43.29&40.33&24.18&-0.32\\
NGC3147&	Sy2	&&41.44&43.68&40.21&20.46&1.23\\
NGC3227&	Sy1.5	&&41.57&42.90&41.23&22.90&0.24\\
NGC3310&	H{\sc ii}/AGN	&&40.32&42.99&39.46&20.85&0.86\\
NGC3486&	Sy2	&&39.93&42.28&38.27&20.28&0.80\\
NGC3516&	Sy1.5	&&43.77&43.23&40.80&23.51&1.63\\
MESSIER066&	Sy2	&&39.48&43.12&39.40&20.32&0.07\\
NGC3690&	H{\sc ii}/AGN	&&41.30&44.39&40.89&20.78&0.41\\
NGC3735&	Sy2	&&40.21&43.45&40.72&20.48&-0.51\\
NGC3976&	Sy2	&&40.09&42.96&39.51&20.04&0.52\\
NGC3982&	Sy2	&&40.16&42.55&39.87&23.34&-0.12\\
NGC4013&	LINER	&&39.02&42.38&37.07&22.95&1.77\\
ARP244&	H{\sc ii}	&&40.39&43.55&39.26&20.51&1.12\\
NGC4051&	Sy1.5	&&40.91&42.60&39.72&23.27&1.07\\
NGC4151&	Sy1.5	&&42.09&43.07&41.48&22.78&0.57\\
NGC4214&	H{\sc ii}	&&38.95&42.21&39.37&20.78&-0.43\\
NGC4253&	Sy1.5	&&42.96&43.51&41.85&20.26&1.11\\
MESSIER099&	H{\sc ii}/AGN	&&40.00&44.12&39.51&20.46&0.49\\
MESSIER100&	LINER	&&40.06&43.59&38.98&20.30&1.07\\
NGC4388&	LINER	&h&42.89&43.61&41.44&23.60&1.08\\
NGC4414&	H{\sc ii}/LINER	&&39.33&42.98&37.38&20.20&1.95\\
NGC4449&	H{\sc ii}	&&38.51&41.57&38.44&20.30&0.05\\
3C273&	Sy1	&&45.82&46.15&-&20.26&-\\
NGC4490&	H{\sc ii}/AGN	&&39.49&42.57&38.32&21.30&1.16\\
MESSIER088&	Sy2	&&41.69&43.87&39.86&24.18&0.33\\
NGC4559&	H{\sc ii}	&&39.91&42.29&37.49&21.04&2.41\\
MESSIER090&	H{\sc ii}/AGN	&&37.25&40.52&37.60&20.45&-0.35\\
MESSIER058&	LINER	&b&41.35&43.23&39.73&20.45&1.62\\
NGC4593&	Sy1	&&42.84&43.33&41.27&20.30&1.56\\
MESSIER104&	LINER	&&38.89&41.10&36.18&20.85&2.71\\
NGC4631&	H{\sc ii}	&&38.90&43.08&37.39&21.32&1.51\\
NGC4666&	H{\sc ii}/LINER	&&40.48&43.71&-&20.23&-\\
NGC4725&		&&39.25&42.31&-&19.95&-\\
UGC08058&	Sy1	&&42.47&44.70&-&22.85&-\\
NGC4968&	Sy2	&&43.22&43.53&41.91&24.48&-1.15\\
NGC5005&	LINER	&b&39.91&42.97&39.03&20.04&0.88\\
MESSIER063&	LINER	&&39.17&42.95&38.19&20.60&0.94\\
MCG-03-34-064&	Sy2	&h&42.95&44.16&42.08&23.60&0.12\\
NGC5170&		&&39.34&42.74&-&21.18&-\\
NGC5194&	Sy2	&&40.43&43.13&40.25&24.18&-0.93\\
ESO383-G035&	Sy1.2	&&42.75&43.04&41.08&20.59&1.64\\
MESSIER083&	H{\sc ii}	&&39.41&43.62&39.29&20.60&0.11\\
IRASF13349+2438&	Sy1	&&43.81&45.66&44.26&22.02&-0.46\\
NGC5256&	Sy2	&&42.21&44.13&-&23.24&-\\
NGC5253&	H{\sc ii}/AGN	&&37.94&41.66&40.47&20.60&-2.53\\
MRK0273&	Sy2	&&42.84&44.26&43.17&23.78&-1.08\\
IC4329A&	Sy1.2	&&43.75&44.21&42.88&20.66&0.87\\
UGC08850&	Sy2	&h&43.15&44.85&-&23.59&-\\
NGC5506&	Sy2	&i&42.85&43.43&41.92&22.51&0.84\\
NGC5548&	Sy1.5	&&43.43&43.86&41.69&20.20&1.74\\
NGC5775&	Sy2	&&39.84&43.50&42.36&21.72&-2.74\\
2MASXJ15115979&	Sy1	&&43.94&44.54&43.26&20.92&0.35\\
VV705&	H{\sc ii}/AGN	&&42.35&44.47&41.84&23.93&-0.19\\
UGC09944&	Sy2	&&41.60&44.04&42.16&20.45&-1.32\\
2MASXJ15504152&	Sy2	&h&43.00&44.10&42.59&24.18&-0.62\\
NGC6240&	LINER	&&43.49&44.34&42.12&24.05&0.30\\
NGC6286&	H{\sc ii}/AGN	&&40.58&43.96&-&20.26&-\\
NGC6552&		&h&42.88&43.95&-&24.18&-\\
AM1925-724&	Sy1	&&42.79&44.77&-&23.58&-\\
NGC6810&	H{\sc ii}/AGN	&&39.91&43.53&40.62&20.70&-0.71\\
NGC6890&	Sy2	&&42.18&43.12&42.02&24.18&-1.41\\
MRK0509&	Sy1.2	&&44.00&44.31&42.48&20.62&1.52\\
ESO286-IG019&	H{\sc ii}/AGN	&&42.31&44.55&-&23.69&-\\
NGC7090&		&&39.82&42.27&-&21.15&-\\
NGC7172&	Sy2	&&42.74&43.26&38.98&22.93&3.57\\
NGC7213&	H{\sc ii}/LINER	&b&42.66&43.08&41.44&20.04&0.77\\
IC5169&	H{\sc ii}/AGN	&&40.72&43.20&-&20.04&-\\
NGC7252&		&&40.36&43.69&-&20.30&-\\
3C445&	Sy1	&&43.94&44.94&43.45&23.54&0.27\\
NGC7314&	Sy1.9	&h&42.33&42.85&39.94&21.78&2.37\\
MCG-03-58-007&	Sy2/LINER	&h&42.74&44.11&41.69&23.44&0.58\\
NGC7469&	Sy1.2	&&43.25&44.37&-&20.64&-\\
NGC7479&	Sy2	&&42.04&43.69&39.87&24.30&0.79\\
ESO148-IG002&	H{\sc ii}/AGN	&&43.21&44.59&42.58&24.18&-0.68\\
NGC7552&	H{\sc ii}	&&40.21&43.79&-&20.11&-\\
NGC7582&	Sy2	&i&42.61&43.56&41.48&24.18&-0.34\\
NGC7674&	Sy2	&h&43.62&44.49&42.08&24.18&-0.01\\
NGC7714&	H{\sc ii}/AGN	&&40.52&43.27&-&21.15&-\\
NGC7771&	H{\sc ii}/AGN	&&40.92&44.05&-&20.70&-\\
MRK0331&	H{\sc ii}/AGN	&&40.67&43.95&41.08&21.08&-0.41\\

\hline
\end{longtable}
\end{landscape}

\twocolumn

%\footnotetext\thanks{This paper has been typeset from a \TeX/\LaTeX file prepared by the author.}

\label{lastpage}
\end{document}